\documentclass[sigconf]{acmart}

\usepackage{booktabs}
\usepackage{multirow}
\usepackage{makecell}

\newcommand{\up}[1]{\textcolor{black}{#1}}

\copyrightyear{2026}
\acmYear{2026}
\setcopyright{cc}
\setcctype{by-nc-nd}
\acmConference[CHIWORK '26]{Proceedings of the 5th Annual Symposium on Human-Computer Interaction for Work}{June 22--25, 2026}{Linz, Austria}
\acmBooktitle{Proceedings of the 5th Annual Symposium on Human-Computer Interaction for Work (CHIWORK '26), June 22--25, 2026, Linz, Austria}
\acmDOI{10.1145/3808045.3808070}
\acmISBN{979-8-4007-2429-9/2026/06}
\usepackage{enumitem}

\begin{document}

\title{Concerns and Strategic Responses of Older Workers Navigating Generative AI in Bridge Employment} 

\author{Aditya Nayak}
\affiliation{%
 \institution{University of Pittsburgh}
 \city{Pittsburgh}
  \state{Pennsylvania}
  \country{USA}}
\email{aditya.nayak@pitt.edu}
\orcid{0009-0009-5517-0932}

\author{Aakash Gautam}
\authornote{Co-senior authors.}
\affiliation{%
 \institution{University of Pittsburgh}
 \city{Pittsburgh}
  \state{Pennsylvania}
  \country{USA}}
\email{aakash@pitt.edu}
\orcid{0000-0001-8023-4648}

\author{Rama Adithya Varanasi}
\authornotemark[1]
\affiliation{%
 \institution{New York University}
 \city{New York City}
  \state{NY}
  \country{USA}}
\email{varanasi.r@nyu.edu}
\orcid{0000-0003-4485-6663}


\begin{abstract}
Generative AI (GenAI) is transforming workplaces at a rapid pace. This disproportionately affects vulnerable communities, including older workers (OWs) who re-enter the workforce through bridge employment prior to final retirement. Through in-depth semi-structured interviews with 21 professionals, we examine how OWs navigate GenAI-driven disruptions while pursuing bridge roles, focusing on their concerns about GenAI integration and their responses to these changes. Our findings show that OWs experienced both temporal and structural disruptions across all stages of the bridge employment decision-making process due to GenAI. In response, they reconfigured their tasks through different forms of boundary work aimed at restoring stability and continuity. We conceptualize these responses as AI resilience, which reshaped OWs' bridge employment decision-making into an ongoing process of negotiation and adaptation. We conclude by offering recommendations to reduce burnout among OWs by balancing individual-level AI resilience strategies with meso-level AI resilience collectives and macro-level adversarial and contestable AI-mediated organizational structures.
\end{abstract}

\ccsdesc[500]{Human-centered computing~Empirical studies in HCI}
\ccsdesc[300]{Human-centered computing~Collaborative and social computing}
\ccsdesc[300]{Social and professional topics~Employment issues}
\keywords{AI resilience, future of work, human-AI interaction, aging workforce, older adults, generative AI}


\maketitle

\section{Introduction}
By 2050, older workers (OWs)\footnote{For this study, we adopt the labor studies definition of older workers as individuals aged 40 and above ~\cite{henretta2018life, EEOC1967}.} are projected to comprise nearly 20\% of the global population~\cite{hedge2012oxford}. During the same period, median life expectancy is expected to rise by 4.5 years, reaching 78.1 years~\cite{vollset2024burden}. These demographic shifts, coupled with insufficient pension coverage and rising financial insecurity in later life, are compelling many OWs to extend their working years  \cite{Alcover_2014}. A growing manifestation of this trend is~\textit{bridge employment}, a form of paid work undertaken after a primary career but before full retirement~\cite{beehr2015working}. 
Bridge employment functions as a buffer mechanism, providing temporal and structural stability between career work and full retirement, while supporting both financial security and continuity in personal and social domains~\cite{oh_bridge_2024, saradha_vathana_selvendran_bridge_2022}.

Technology can be a key enabler in navigating bridge work with opportunities such as digital platforms and remote work that can expand access to flexible, modular forms of employment that better align with older workers’ needs~\cite{massihzadegan_preparing_2022, sharit_employability_2009}. 
At the same time, rapid technological innovation, particularly the rise of generative AI (GenAI), is reshaping workplaces in ways that differ from prior technological shifts. 
As a general-purpose technology with low barriers to adoption, GenAI simultaneously affects the skills that employers value, work roles, and the dynamics between workers and employers  \cite{Law2025, aisa_automation_2023}.
For OWs pursuing bridge employment, these shifts raise a critical question of how well they can respond to GenAI and navigate GenAI-mediated workplaces. 
Our study responds to this gap by asking: (\textbf{RQ-1}) \textit{What are the key concerns of older workers (OWs) regarding the proliferation of generative AI across different aspects of their bridge employment}?, and (\textbf{RQ-2}) \textit{How are older workers (OWs) responding to these concerns}?


To answer our research questions, we conducted an exploratory study comprising semi-structured interviews with 21 OWs (avg. age = 57 years) who were either employed in bridge work or were transitioning into such roles across diverse fields of knowledge and creative work, a context where declining retirement protections and rapid GenAI adoption create particularly acute pressures on bridge employment (see Section \ref{sec:lit_india}).
OWs shared multiple concerns related to GenAI across all stages of their bridge employment decision-making process (see Figure \ref{fig:be-results}). GenAI disrupted OWs’ planning and intention to pursue bridge employment by introducing uncertainty about the relevance of their skills and by destabilizing established trajectories of bridge roles. For OWs already engaged in bridge employment, GenAI raised concerns about deprofessionalization by eroding established work pathways and positioning them in competition with younger workers. Bridge employment is a critical mechanism through which OWs maintain economic and psychological stability following the departure from their career job. Our findings show that GenAI undermined this stability by introducing both temporal and structural disruptions. 
Unlike prior technological shifts that affected discrete stages of work transitions, GenAI disrupted the entire decision-making process,  altering the nature of bridge employment from a planned pathway to an uncertain endeavor that required ongoing negotiations and adaptations.

At the same time, consistent with recent HCI research on OWs (e.g., \cite{Zhao_2023, Jelen2023, Kope2018, Convertino_2007, kobayashi_motivating_2015, brewer_why_2016}), we observed that OWs adopted proactive strategies to mitigate these concerns and restore stability in their work. 
During the planning and intention phase, OWs reoriented their skills and professional identities to complement GenAI while drawing on qualities they perceived as strengths associated with their seniority. 
Among OWs who had undertaken bridge employment, participants reconfigured their work by engaging in boundary work, deliberately crafting and maintaining boundaries to demarcate, market, and protect their contributions while selectively leveraging GenAI’s capabilities. 
This boundary work even enabled some OWs to pursue novel opportunities and take risks that they would typically avoid in bridge employment contexts. 
We conceptualize these responses through the lens of \textit{AI resilience}, which we see as the capacity of workers to recognize, absorb, and strategically respond to AI-driven disruptions by drawing on accumulated professional experience to restore stability and continuity in their work.
AI resilience, in this sense, captures the ongoing, iterative process through which workers negotiate their relationship with AI systems in their work.
We provide evidence of AI resilience by showing how OWs engaged in a delicate balance of \textit{observing} and \textit{responding}, allowing them to engage tactfully with GenAI without becoming over-reliant on it. This stands in contrast to challenges documented among younger workers in human-GenAI collaboration, where over-reliance has been shown to undermine effective engagement in work contexts \cite{Lee_2025, shukla_-skilling_2025}. Building on this perspective, we propose re-imagining human-AI interactions through the lens of AI resilience and adversarial design. Our study makes the following contributions:

\begin{itemize}
\item We identify distinct temporal and structural disruptions that OWs experience across the bridge employment decision-making process as GenAI is integrated into bridge roles.
\item We show how OWs respond to these disruptions by reorienting their roles and reconfiguring their bridge tasks through boundary work, enabling them to restore and maintain work stability.
\item We extend bridge employment theory by demonstrating how GenAI transforms the decision-making process from a linear transition into continuous negotiation and adaptation.  
\item We introduce AI resilience as a framework for understanding how older workers leverage accumulated professional experience to strategically adapt to GenAI in their work, and we outline meso- and macro-level pathways for distributing the burden that individual resilience practices impose.
\end{itemize}

\section{Related Work}
In this section, we review prior research that motivates our research questions. We begin with a brief introduction to bridge employment and the associated decision-making process framework, which articulates the steps OWs take in pursuing bridge employment. We then review prior HCI research in the contexts of OWs and bridge employment.

\subsection{Bridge Employment} \label{lit-bridge}
Bridge employment refers to paid work undertaken after retirement from a primary career job but before complete withdrawal from the labor force~\cite{Zhan2009}. Originating in organizational studies, the term ``bridge'' describes the period between leaving one’s main career and entering full retirement. Although the concept has only recently been formalized, related ideas have been explored for much longer~\cite{doeringer1990business}. Traditionally, the transition to bridge employment occurred later in life, often around the conventional retirement age of 65 (or its global equivalent). However, this shift is happening earlier, sometimes as early as age of 40~\cite{Peng_Chan_2019}, due to changing retirement patterns, with more individuals choosing early, informal, partial, or phased retirement~\cite{Kantarci2008}. 

The transition itself is a complex multi-stage decision-making process. \citet{beehr2015working} modeled this decision process as four steps (see Figure \ref{fig:be-results}). The first stage is \textbf{planning}, in which senior professionals anticipate retirement from their primary career and consider ways to continue paid (bridge) work, evaluating available options, constraints, and risks. The second stage is \textbf{intention}, where individuals assess the options identified during planning and form concrete intentions about whether to pursue a particular bridge role. The third involves \textbf{behavior} during which individuals enact these intentions by entering bridge employment and taking up specific roles and work arrangements. The fourth stage is called \textbf{adjustment}, in which individuals evaluate their experiences in bridge employment, assessing satisfaction, fit, and well-being, which informs decisions to continue, modify, or exit bridge work. While \citet{beehr2015working} do not portray the decision-making process as strictly linear, their model nonetheless reflects a broader tendency in bridge employment research to conceptualize the transition through a temporal lens, treating it as a time-ordered transition rather than a static work arrangement.


Bridge employment itself can also take multiple forms. Some workers remain in their career field but take on a different role or reduced responsibilities, while others shift to opportunities outside their previous line of work~\cite{gobeski2009retirees}. These are often referred to as career-consistent and non-career bridge employment \cite[Ch.~1]{Alcover_2014}. 
Non-career bridge jobs are a viable option for older workers as they offer greater flexibility in exchange for reduced compensation ~\cite{Shultz_2003}. The timing of entry into bridge employment can also vary. Some individuals transition directly into bridge work upon leaving their primary careers, while others do so after a period away from the workforce~\cite{Wang2008}. Bridge employment can further differ by work arrangement. Individuals may be self-employed or work for another organization, in steady positions or through intermittent gig opportunities~\cite{beehr2015working}.


These different forms of bridge employment are shaped by motivational factors that operate at multiple levels: micro-level factors relating to individual circumstances such as financial status and health~\cite{kim2000working,  Wang2008}; meso-level factors including family and workplace  context~\cite{shultz2011psychological}; and macro-level factors involving organizational policies, labor market trends, and the regional  economy~\cite{Nakai_2011}. This multi-level structure becomes relevant when considering how workers respond to disruptions, as adaptive strategies may operate at all of these levels.

Taking up bridge work beyond career responsibilities entails several benefits, most significant among them is enhancement of occupational well-being and life satisfaction~\cite{Drentea_2002}. 
In bridge work environments, where individuals work shorter hours and have reduced responsibility, the appropriate level of mental engagement can improve mental health \cite{saradha_vathana_selvendran_bridge_2022}. 
Beyond individual benefits, workers' engagement with bridge work also benefits governments to maintain their labor, by reducing negative outcomes and dysfunction in social security contributions~\cite{ekerdt1990defining}. 
However, benefits are strongly dependent on motivations, with stronger financial reasons for joining bridge work being associated with lower satisfaction~\cite{Dingemans_Henkens_2014}. OWs who enter bridge employment after retirement often encounter fewer opportunities and greater challenges~\cite{MarshMcLennan2018, Topa_2014}. For example, they can experience greater inter-personal conflicts with younger employees due to stark differences in work identities~\cite{Ho_Yeung_2021}. 
The proliferation of GenAI in work adds a layer of complexity to this context.
However, much of the early bridge employment scholarship focuses on individuals' motivations and outcomes.
There remains a notable gap in our understanding of how workplace technologies, and GenAI in particular, shape older workers’ access to and experiences within bridge work.

\subsection{Bridge Employment in HCI}
Within Human--Computer Interaction (HCI) research, OWs have long been a central focus~\cite{brewer_why_2016}. Early work primarily examined their abilities and challenges under the broader umbrella of older adults and their use of information technologies~\cite{gell_patterns_2015}, often adopting a deficit-oriented perspective that emphasized limitations in skills or cognitive capacity and portrayed them as less efficient, more error-prone, and in need of simplified interfaces or training~\cite{niehaves_internet_2014, vines_age-old_2015}. In contrast, more recent scholarship has critiqued this framing, advocating for inclusive and empowerment-oriented approaches that foreground agency, participation, and design for diversity~\cite{righi2017we, vines2012cheque, rogers2014never}. Older individuals tend to have a positive attitude toward technology adoption~\cite{czaja_impact_2007, lee_age_2011, pang_technology_2021} and show a willingness to learn and use technology if they can understand the potential benefits associated with it~\cite{barnard_learning_2013, lee_perspective_2015}.

Within this shift, there has been renewed attention to the existing skills and practices of older individuals, particularly those that foster social interaction, inclusion, and self-expression as valuable starting points for technology design ~\cite{lindsay2012empathy, kuoppamaki2021designing, vines2012questionable}. Analysis of blog posts by older individuals on their encounters with ageist stereotypes~\cite{lazar_going_2017} showed that seniors used blogging as a space to share ``concealed stories''~\cite{bell_storytelling_2020} of their experiences to develop a collective narrative against ageism as a networked-public~\cite{disalvo_design_2009}. More recently, in the context of emerging AI-driven systems, voice-first technologies have shown promise by leveraging older individuals' conversational and storytelling abilities, thereby enabling more natural and accessible forms of interaction~\cite{cuadra23, Pradhan2020}. Much of this HCI research, however, has focused on older adults in the context of their personal lives and everyday activities.

Emerging research in organizational studies shows growing evidence that rapid technological advancement exacerbates the vulnerabilities OWs face when they re-enter the workforce~\cite{MarshMcLennan2018}. One challenge is the abrupt introduction of emergent technologies in bridge employment roles. These may be tools they had limited engagement in their primary careers and for which they have received little to no scaffolding or support~\cite{fletcher-watson_strategies_2016}. As a consequence, such workplace technologies can exacerbate existing issues such as ageism~\cite{Butler_2010, levy_mind_2003}, misconduct, and conflicts while working with the younger population, contributing to detrimental effects~\cite{coudin_help_2010, levy_stereotype_2009}. Despite these issues, OWs do embrace and rely on workplace technologies~\cite{Pak2017}. When deliberately designed, technology-enabled work such as crowd work can better align with older workers' motivations for working, including cognitive stimulation and work-related identities~\cite{kobayashi_motivating_2015}. These new forms of work can also bring much-needed flexibility in the working dynamic, providing both temporal and spatial autonomy~\cite{Atkinson2016}.

While there is limited research in HCI that focuses exclusively on OWs and technology-mediated labor in the workplace, emerging studies are recognizing the opportunities and challenges these new forms of technology bring. A growing area of interest explores how to engage OWs’ knowledge and abilities to enhance informal work~\cite{Zhao_2023}. For example, researchers have examined ways to leverage OWs’ subject expertise in participatory activities such as making~\cite{Jelen2023} and hackathons~\cite{Kope2018} to improve both the diversity and overall outcomes of co-design practices. Within formal work contexts, an early area of research was the effort to understand and improve intergenerational working dynamics around technology-mediated working environments~\cite{Convertino_2007}. More recent studies have investigated new forms of gig work to better understand the challenges aging workers face as they take on these roles~\cite{kobayashi_motivating_2015, brewer_why_2016}. Despite these advances, HCI research has paid limited attention to how older workers’ need for bridge employment shapes their experiences in technology-mediated work. Our study fills this gap by addressing the underexplored experiences of older workers transitioning to bridge employment in the current age of AI.

\subsection{Resilience in AI Contexts}  
Resilience, broadly defined as the capacity to cope with and recover from adversity, has a long history in gerontology research \cite{lima_resilience_2023, ye_psychological_2024, cosarderelioglu_frailty_2025}. 
Studies of older adults have characterized resilience as a proactive process through which individuals draw on accumulated personal resources to overcome challenges and restore stability \cite{van2013ability, Wiles2012}. 
In the context of socio-technical systems,~\citet{Heeks2019} extended this concept to  \textit{technological resilience}, emphasizing individuals' capacity to cope with and thrive in response to external socio-technical shocks. 
More recently, ~\citet{glassman2024ai} proposed the notion of \textit{AI-resilient interfaces}, arguing that systems should be designed to support users in noticing AI actions and judging their implications, so that individuals can protect themselves against potential harms.  

These frameworks offer a useful starting point but remain limited in two respects. 
First, they are primarily interface-level concepts focused on individual interactions with specific AI systems, whereas the disruptions GenAI introduces to bridge employment operate across work practices,  professional identities, and career trajectories, all simultaneously. 
Second, existing resilience scholarship has examined responses to life transitions, such as economic challenges, health decline, or bereavement ~\cite{van2013ability, vyas2017everyday}, but has not addressed technology-driven disruptions to work. 
We extend these scholarships by examining how older workers enact resilience in response to GenAI-driven disruptions across their bridge employment, and by formalizing the mechanisms through which this resilience operates.

\subsection{Retirement and Aging: Indian Context}
\label{sec:lit_india}

\up{
India presents a particularly relevant context for the demographic, economic, and technological forces shaping bridge employment. 
The working age population is expected to peak between 2030 and 2035, while the proportion of older adults has been rising since 2010 and is projected to reach 15\% of the population within the next decade ~\cite{jain_population_2025}.
A 2024 national survey found that only 7\% of elderly respondents were willing to work, while 50\% felt too old for a job and 31\% felt they lacked the required skills to work ~\cite{helpage_india_ageing_2024}. 
Despite this, 24\% reported continuing to participate in the workforce, which suggests the structural pressures that compel labor workforce engagement. 
}

\up{
From scholarship and popular media, we note several pressures for continued workforce participation. 
Primarily, there is a progressive erosion of institutional protections that historically buffered older Indians during retirement ~\cite{iyr_hidden_2025, chakrabarty_union_2026, money_control_low_2025, sun_life_news_2026, guan_research_2025, bungsut_age_2024}. 
This can be seen, for example, in the fact that employer-guaranteed pensions have shifted toward employee-managed schemes across multiple sectors ~\cite{path_changing_2024, berger_defined_2012}, while public healthcare, pensions, and insurance coverage remain limited ~\cite{sahoo_charting_2024}.
Close to two-thirds (65\%) of elderly respondents in the same national survey reported feeling financial insecurity, and only 29\% reported access to social security schemes or pensions ~\cite{helpage_india_ageing_2024}.
The joint family system, which historically functioned as a de facto social safety net, has weakened as urbanization-driven youth migration has given rise to nuclear family structures  ~\cite{dutta_elderly_2026}. 
Roughly 20\% elderly Indians live alone or with a spouse ~\cite{mahambare_india_2026} and frequently experience economic instability, social isolation, and health vulnerability.
}

\up{
At the same time, India has become one of the largest markets for GenAI deployment. 
AI adoption in India's service sector stands at 30\%, which exceeds the global average of 26\% ~\cite{rajmohan_impact_2025, indiaai_india_2026}. 
This adoption is driving major restructuring across knowledge and creative work roles ~\cite{inamdar_tcs_2025, rajmohan_impact_2025}.
A 2024 survey of India's white-collar workforce found that 68\% expected AI to partially or fully automate their jobs within five years; 40\% perceived that their current skills would become redundant ~\cite{chakrabarti_labour-force_2024}. 
India is also home to one of the largest workforces engaged in knowledge and creative work within global markets, with occupational roles that are broadly generalizable to other labor contexts ~\cite{ilo2024}.
The simultaneous interplay of these forces, particularly of declining retirement protections pushing older workers toward bridge employment, and rapid GenAI adoption restructuring the very roles they seek, makes the Indian context a particularly productive space to examine how older workers navigate 
GenAI-driven disruptions in bridge employment.
}

\section{Methodology}
To answer our research questions, we conducted an exploratory qualitative study comprising in-depth semi-structured interviews with older workers (OWs) in India (avg. age = 57 years). Through these interviews, our goal was to: (1) understand how OWs perceived GenAI proliferation in their bridge employment and (2) investigate how they adapted GenAI based on their work experiences and the challenges they encountered. The research was approved by the IRB and took place between January and March of 2025. 

\begin{table*}
 \center
 \renewcommand\arraystretch{1.6}
 \footnotesize
 \begin{tabular}[t]{|p{1.5in}|p{2.43in}|}
 \hline
 \multicolumn{2}{|l|}{{\bf Total Participants} (n=21)}\\ 
 \hline
  Gender & 
       \begin{tabular}{lll}
           Women:11  & Men: 10 \\
       \end{tabular}\\
\hline
 Age (years) & 
       \begin{tabular}{llll}
         Min: 47 & Max: 71 & Avg: 57 & S.D: 7.8\\
       \end{tabular}\\
\hline
Role  & 
       \begin{tabular}{p{2.3in}}
           Content writer (x3); Product manager (x2); Voice actor; Journalist (x2); Film maker; Program manager (x2); Lawyer (x3); Paralegal; Creative manager; Financial manager (x3); 
           Faculty (x2) \\
       \end{tabular}\\
\hline

Bridge employment type & 
       \begin{tabular}{lll}
           Employed:8; & Contract/consultant:3 & Self-employed:10  \\
       \end{tabular}\\
\hline
Bridge employment experience (years)  & 
       \begin{tabular}{llll}
            Min: 1 & Max: 7 & Avg: 3 & S.D: 2.1\\
       \end{tabular}\\
\hline
Career experience (years) & 
       \begin{tabular}{llll}
           Min: 10  & Max: 45  & Avg: 28.2  & S.D: 10.4 \\
       \end{tabular}\\
\hline
GAI tools & 
       \begin{tabular}{p{2.3in}}
          ChatGPT (x19); CoPilot (x4); Midjourney (x2); Perplexity (x2); Gemini (x6); MurphAI (x1); Eleven labs (x1); Grok (x2); Sora (x2); Dalle (x2); AlliAI (x1); Deepseek (x1);  Claude (x3); Custom-GenAI (x2)\\
       \end{tabular}\\

\hline
\end{tabular}
\caption{Demographic details of older workers interviewed for the study.}
 \label{tab:org-participants}

\end{table*}

\subsection{Participant Recruitment}
We used a combination of purposive and snowball sampling strategies to recruit older working professionals in India who were either working in bridge employment roles or in the process of transitioning to such roles across various knowledge and creative industries. 

We initiated our interview study by sharing our recruitment message on various social networks and work platforms that are prominent hubs for knowledge and creative work, such as LinkedIn, Upwork, Fiverr, Reddit, and Twitter (X), along with a screening survey that captured their (1) age, (2) stage of career, (3) views around retirement, and (4) use of GenAI tools in their work. From the data of the individuals who completed the survey and showed interest in the study, we shortlisted 21 candidates to conduct the interview. During the shortlisting process, we looked for diverse participants around bridge employment, including (1) their perceptions of retirement, (2) the inclination to work after retirement, and (3) the type of GenAI tools they were using. We also ensured that the shortlisted participants represented diverse professional roles that are broadly reflective of the global workforce and had at least two years of exposure to GenAI tools. 

\subsection{Participant Background and Demographics}
Out of 21 participants, 10 identified themselves as men and 11 as women. The average age of the participants was 57 years (min = 47 years, max = 71 years). \up{Although a mean age of 57 may seem relatively young for aging research, occupational literature defines older workers as those aged 40 and above. This broader threshold reflects the considerable variability in early retirement timing and recognizes that bridge employment, the transitional work undertaken between early and full retirement, is most relevant during the years immediately following early retirement \cite{EEOC1967}}. Participants had various roles within knowledge and creative work, including journalist, content writer, lawyer, academic faculty, IT and product managers, voice actor, and graphic designer. These participants worked in diverse industries, such as finance, business, social media, law, entertainment, creative fiction, technology development, and academia. Seven participants were in the process of transitioning to bridge employment. Of the remaining fourteen, six had taken early retirement for various reasons before moving into bridge employment, while eight began bridge employment after retiring from their career jobs. On average, participants earned a monthly salary of Rs. 1 lakh (one hundred thousand; approximately USD\$~1,100), representing upper-middle-income white-collar professionals with earnings comparable to similar professional roles across global labor markets. Participants had an average of two years of experience using GenAI tools in both work and personal contexts. The most commonly used tool was ChatGPT (n=19), followed by Gemini (n=6). \up{Although participants used a variety of GenAI tools across work and personal contexts, including text-based systems (e.g., ChatGPT, Gemini, Perplexity),  image and video generators (e.g., Midjourney, DALL-E, Sora), and voice synthesis tools (e.g., MurphAI, ElevenLabs), our analysis centers on the shared experience of navigating AI-generated outputs in professional work rather than on the affordances of individual tools.}

\subsection{Procedure}
We conducted our semi-structured interviews remotely through video calls. Before the study session, we contacted the participants to inform them about the study objectives, obtain informed consent, and set realistic expectations, such as the voluntary nature of the study. This included familiarizing participants with the interview study procedure and clarifying our neutral stance towards GenAI tools. We also obtained participants' orientation and usage of GenAI tools to customize our study to their perceptions and usage behaviors. Participants were compensated with a gift card of an approximate value of \$15. The interviews lasted anywhere between 1 to 1.5 hours (avg. = 1.15 hours) and were conducted in English. 

To manage the scope of our research study, our questions focused exclusively on the sociotechnical impacts of GenAI on their bridge work. The protocol consisted of four sections, wherein we captured (1) their bridge work practices (e.g., ``\textit{Can you walk me through a typical day of your work? How is it different to the work you did before retirement?}''), (2)  their perception of GenAI proliferation in bridge work (e.g.,``\textit{Describe a specific moment when you first encountered GenAI in your work context.}''), (3) benefits and challenges around GenAI in their bridge employment (e.g., ``\textit{Describe a concrete situation where using GenAI created a challenge or tension in your bridge work.}''), and (4) how are they responding to the challenges (e.g., ``\textit{Can you walk me through a routine in your current job that you changed in response to GenAI?}"). Multiple authors conducted the interviews and captured detailed notes, such as participants' concerns, their work activities, and different GenAI tools they mentioned. The notes were used as memos in the analysis of the study.

\subsection{Data Collection and Analysis}
Our study resulted in 23.4 hours of audio-recorded interviews. In addition, we also captured several pages of notes during the interviews. Audio data was transcribed verbatim. All of the aforementioned data was analyzed using inductive thematic analysis. As a first step, we started taking multiple passes of our transcribed data to familiarize with the participant's personal accounts. This exercise was started in parallel with the interview process, allowing us to refine the interview protocol \cite{smith2024qual}. Subsequently, we conducted multiple rounds of open coding while keeping aside our preconceived notions and theoretical assumptions \cite{Braun_Clarke_2006}. During this phase, we used memos to capture our reflections. The overall process resulted in  88 codes. During this process, we prioritized insights that were directly relevant to generalizable sociotechnical practices of bridge employment. Through several rounds of discussions among the authors, we removed overlapping and duplicate codes. The resulting codebook consisted of 61 codes. Examples included \textit{Attitudes towards learning}, \textit{Strategies to overcome challenges}, and \textit{Guidelines and directives}. 

In the second step, we adopted an interpretive abductive approach \cite{Soden2024} using focused coding \cite{Hennink_2019} to examine relationships across codes and identify higher-level themes relevant to our research questions. To achieve this, we employed \citet{beehr2015working}'s bridge employment decision-making process, comprising \textit{planning}, \textit{intention}, \textit{behavior}, and \textit{adjustment}, as an analytical framework.
As part of this approach, we engaged in prolonged interaction with the data and peer debriefing to resolve analytic disagreements and reach thematic saturation \cite{Creswell2000}. The resulting themes are organized into two overarching categories: (a) GenAI concerns and challenges, and (b) proactive changes undertaken in response to GenAI, both situated within stages of the bridge employment decision-making process. Examples of the resulting themes include \textit{Uncertainties in transition} , \textit{flexible boundaries}, and \textit{undertaking increased risk}. Based on this final structure, we present our findings below.

\subsection{Positionality}
All authors were born and raised in the Global South, and their personal experiences with older workers have shaped and informed this study. While this background enabled a nuanced investigation, we also made a deliberate effort to identify and mitigate our own biases. 
This included acknowledging that we are based in computing programs currently, and the second and third authors were trained in computer science. Through discussions and critical reflections, we tried to be more aware of our perspectives about technology as we designed the interviews, analyzed the data, and wrote this manuscript. We also ensured participants had the flexibility to reschedule, cancel, or conduct the interview in multiple sessions to accommodate their personal needs. Drawing on feminist methodologies \cite{Trier2012}, we explicitly stated our positionality and key assumptions during the interviews. Given the remote nature of the interviews, we embedded verbal and non-verbal actions to make participants comfortable and fostered conversation where the participants shaped the direction of the conversations.  
We also recognize that our participants, while diverse in professional backgrounds, represent upper-middle-income white-collar professionals whose experiences may not reflect those of older workers in informal or lower-income sectors.

\section{Findings}
We organize our findings using Beehr and Bennett's \cite{beehr2015working} four-stage bridge employment decision-making framework: planning (anticipating retirement and considering bridge work options), intention (forming concrete plans about specific bridge roles), behavior (enacting bridge employment), and adjustment (evaluating and modifying bridge work practices). We begin by describing OWs’ concerns about GenAI in bridge employment, followed by the proactive responses they undertook to address these concerns.

\subsection{OWs' Concerns About GenAI in Bridge Employment}

\subsubsection{Uncertainty in Bridge Employment Transition: Planning and Intention}
\label{subsec:concern-planning}
OWs engage in bridge employment as part of a planned and gradual transition from full-time work \cite{Wang2008}. The first two stages of this decision-making process (i.e., planning and intention; see section \ref{lit-bridge}) involve careful deliberation to ensure a successful and stable transition while minimizing risk. Our findings show that, compared to previous waves of technological advancement in the workplace, OWs found the pace and scale at which GenAI was being adopted in bridge roles ambiguous and unsettling, impacting the stability they once had in their transition.

\begin{figure*}
    \centering
    \includegraphics[width=1\linewidth]{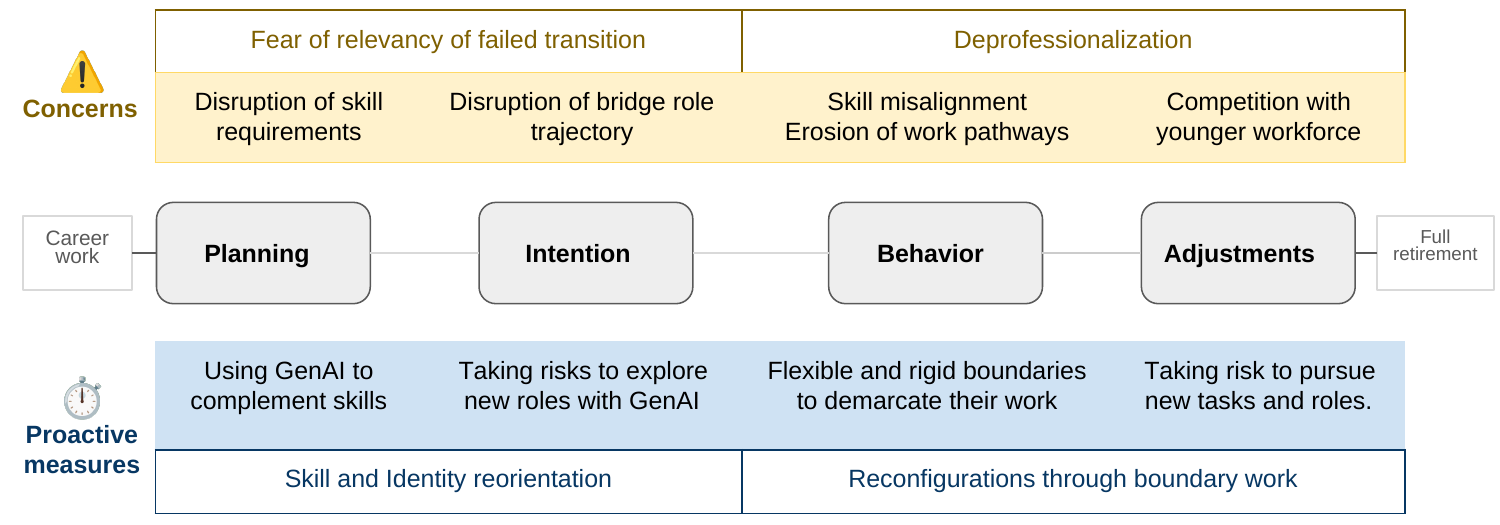}
    \caption{Overview of the study findings mapped onto stages of the bridge employment decision-making process. Findings shown above the decision-making stages in yellow represent OWs' concerns, while findings shown below the stages in blue represent proactive measures taken by OWs.}
    \label{fig:be-results}
\end{figure*}

For instance, participants described how the implementation of GenAI in their prospective bridge roles lacked clear guidelines regarding its adoption and use, as well as clarity about its potential consequences for their bridge work, particularly at such a late stage of their working lives. As a result, OWs struggled to visualize how these anticipated roles might change, even when they were working in similar career jobs at the time. This GenAI-induced uncertainty generated fear and insecurity during the intention phase (see Figure~\ref{fig:be-results}), undermining commitment to particular bridge roles, which is a key outcome of this phase, and affecting the time required for the transition. P6, who was transitioning to a senior editor role as part of the bridge role at the same national newspaper where he also worked during his career role, shared the sense of uncertainty and insecurity arising in the journalistic fraternity because of GenAI in his work:
\begin{quote}
    \textit{``It is disruption because, in my career, I have not seen any change happening at this rapid pace. New applications were introduced from time to time, but that was a continuous process \dots no urgency. This time, things are being pushed blazing fast. Even the top management fears that they may be left behind because they don't know [this technology] \dots We (OWs) are being asked to go to workshops without telling  why \dots There is a general pessimism within the journalistic fraternity \dots insecurity because of [this] rapidly changing technology.''}
\end{quote}

Compared to prior technological transitions, the absence of clear guidelines and protocols for GenAI integration also disrupted OWs’ planning phase by limiting their ability to anticipate how existing skills would remain relevant. This challenge was particularly pronounced for OWs pursuing bridge roles that differed from their career roles, as they struggled to translate their accumulated expertise and current responsibilities into positions reshaped by GenAI. P13, an architect transitioning to academia, was concerned that techniques and skills were rapidly changing in her field, and that her existing skills may not be needed in her bridge role.
\begin{quote}
    \textit{``My knowledge of the kind that I have studied  might not be required \dots The construction techniques have changed, the skills have changed \dots and I don't have any protocol to learn about [GenAI] skills I need to teach to the students. ''}
\end{quote}

\subsubsection{Signals of De-professionalization: Behaviors and Adjustments} \label{subsec:concern-behavior}
Once OWs transition into bridge roles, they often rely on extensive experience and tacit knowledge developed during their career work as core strengths \cite{feldman1994decision}. In our study, OWs described enacting these strengths through everyday behaviors, such as effective decision-making that prioritized quality and reliability. However, with the proliferation of GenAI in their work, OWs were concerned that such practices were increasingly misaligned with emerging expectations that prioritized speed and volume delivered with GenAI. This concern was especially prevalent in OWs who were freelancers or consultants as part of their bridge roles. P8, a creative writer turned freelancer as part of his bridge work, shared how demands shifted from quality to volume:
\begin{quote}
    \textit{``I was stuck with a client who did not want quality work; was willing to pay higher for volume. And I really didn't mind things being rehashed \dots I knew that at some point of time this could be a problem \dots not so soon. When [Gen]AI came along, he said: `I am getting this done for free with [Gen]AI. I don't care about the quality. Why should I pay you so much for the work?''}
\end{quote}

OWs felt that clients expected them to use GenAI in their work while focusing on producing more volume and at faster speeds, making their strengths grounded in deliberation and quality less competitive. 
Beyond individual client relationships, OWs in established professional fields expressed concern that GenAI was eroding entire work pathways dependent on tacit knowledge and professional judgment. For instance, P13 shared her fears on how GenAI was facilitating the circumvention of foundational practices, such as hand-drawn illustration, that have defined the architecture field for a long time and in which they spent years developing expertise:
\begin{quote}
    \textit{``So, the GenAI threat is quite endangering right now. I was with one of the faculty today. He just wrote a simple prompt and generated a complete architectural drawing, just like that! Where is the whole process of sketching it with a hand, like we used to? Are these things going away? It is quite worrying because we learnt all these fundamental aspects as a core part of the work identity. It is a threat to all this.''}
\end{quote}
P13's concern illustrates the disruption in practices, compounded by the absence of a learning protocol, which meant that she could neither anticipate which of her existing skills would remain relevant nor identify a clear pathway for acquiring necessary competencies for future work. This uncertainty made it difficult, adding stress during an already vulnerable transition period. 

Indeed, OWs felt particularly vulnerable and devalued with the proliferation of GenAI in their work, noting that younger workers were more adept at embracing GenAI and developing workflows oriented toward greater speed and higher output. OWs who considered shifting toward productivity-focused GenAI workflows felt that doing so could further jeopardize their positions, as they would be competing directly with younger counterparts. P8 described how job requests in his field were increasingly shifting toward GenAI-assisted work, expressing skepticism about his ability to sustain freelance writing if he continued it as part of his bridge role:

\begin{quote}
    \textit{``Because of this sudden rush of supply [due to GenAI], the rates are down. If I am not going to get a bare minimum per month, is it worth doing this work? \dots When I used to work earlier, before this AI thing came on, I would get something like 4 cents, 5 cents a word on an average \dots Today, [younger] guys on the platform want to work at 1 cent a word. Then I have to think \dots is it worth doing then?''}
\end{quote}

 Several participants reflected on their recent reduction in workloads. They inferred that the scope of their bridge work might similarly shrink, leaving their bridge roles vulnerable to deprofessionalization and eventual replacement. For example, P12, who is a senior content strategist whose core responsibilities include designing banners and SMS content for the marketing team, explained that after his organization introduced Merlin AI (a GenAI tool) into its workflows, his monthly workload gradually decreased, dropping from 1000 pieces to 650 pieces. He felt that this decline was due to some of his work being fed to the GenAI tool by the management. 
 \up{We see this as a form of incremental displacement where GenAI gradually absorbed portions of work and narrowed the scope of bridge employment.}

\subsection{OWs' Responses to GenAI in Bridge Employment}

In light of the concerns described above, OWs demonstrated strategic agency by reorienting their skills and professional identity to position themselves as complementary to GenAI practices. In parallel, they engaged in boundary work where they demarcated where their expertise added value, distinct from GenAI's capabilities. These responses enabled OWs to restore stability while maintaining professional identity and, in some cases, pursuing broader opportunities. 

\subsubsection{Reorienting Skills and Work Identity: Planning and Intention} \label{subsec:response-planning}
Amid the uncertainty introduced by GenAI’s disruption of their targeted bridge roles, we observed OWs reframing their perceptions of GenAI during the planning phase, from a disruptive technology to a complementary one that could help secure bridge employment. OWs described their need to ``\textit{not run away}'' but to ``\textit{rather understand}'' and ``\textit{know how to use [Gen]AI in the work}''. In doing so, OWs often explicitly contrasted their own experiential breadth with the technological fluency of younger workers, viewing GenAI as a means to bridge this gap. For instance, P3, a researcher turned technical writer in her bridge role, shared:
\begin{quote}
    \textit{``We can get along with them [younger colleagues]. I may not be able to do very detailed work [with technology] like them \dots But given an opportunity, we learn [GenAI] and fit into these bridge job \dots this is what the experience gives you \dots an understanding of a broad perspective. I can learn and upgrade myself. And being a senior professional, your decision-making is much better than your juniors.''}
\end{quote}

OWs who planned their transition to roles aligned with their prior careers described engaging in more mindful interactions with GenAI tools within their domains, actively exploring ways to reposition their tacit skills and experiential knowledge as complementary to GenAI’s capabilities. For example, P7, an illustrator who recently transitioned into a storyboarding role for AI-assisted filmmaking, articulated this perspective by framing his illustration expertise as a complementary skill that could support AI-mediated creative processes:

\begin{quote}
    \textit{``[With GenAI] there is no need for someone to be an illustrator to enter the industry. But people who have developed that talent, we have an edge \dots we have the right perception \dots we can scribble and ideate something quickly. We can also draw in a paper, just scan it and upload it into [Gen]AI and it will generate a 3D render \dots we bring that mindset of an illustrator \dots imagining, and arranging the scenes. That is a strength \dots But if someone is getting illustrations also from ChatGPT,  it will give you story and scenarios but they need skill and effort and customize it according to the story \dots that is time-consuming.''}
\end{quote}

P7 described how he reoriented and presented his experience in traditional illustration as a skill that could expedite the overall AI-assisted filmmaking process in his new role.
Both accounts also hinted at the value of their experiential breadth, professional judgment, and tacit skills that GenAI could not replicate or replace.

For OWs who were not interested in transitioning into bridge roles aligned with their prior careers, learning GenAI provided increased confidence and a greater tolerance for risk when exploring new types of work. Participants in this group shaped their identity by describing themselves as ``AI Managers,'' delegating structured technical tasks associated with their new bridge roles while leveraging their accumulated decision-making experience to remain effective. In their view, GenAI was better suited for narrowly defined and structured roles, while their human efforts could focus on handling complex and multifaceted activities. Examples given for the latter included contextualizing ideas, fulfilling cross-role dependencies, and taking leadership and management roles. P17, a product manager deliberating whether to start her own company as part of bridge employment, envisioned GenAI agents as well-suited for structured tasks such as customer support, while senior professionals like her could remain better positioned to stay fluid and manage multiple cross-functional roles:

\begin{quote}
    \textit{``I don't want a single [ChatGPT] tool. I need multiple combinations of such AI agents. Maybe today they will do support, tomorrow they will do testing, the next day they will give a demo. Currently, one person can play multiple roles in the organization. They do not have any fixed type of work, at least in our [product development] context \dots Humans also have constraints on the work timing. For support-type of works, I can use GenAI agents to provide 24x7 support. Once I figure this out, I will go into a more Operator model. ''}
\end{quote}

P17 described this sentiment while evaluating ChatGPT’s new agent feature as part of her efforts to materialize a bridge transition into running her own startup. She shared that her intention to transition and establish a startup strengthened after spending time learning how GenAI agents work and how they could be integrated to handle concrete tasks in her new venture. By delegating activities such as customer service and lead generation to GenAI agents, she envisioned saving time in the early stages to focus on raising capital, instead of hiring and training employees. 
This reorientation does not see GenAI as a threat. Instead, OWs see a possibility to position themselves as orchestrators who can identify tasks that can be delegated to GenAI and those that require human involvement. 
These examples of proactive reorientation of work identity and skill use among OWs contrast with patterns reported in previous research on worker perceptions around GenAI, in which perceived fear and uncertainty surrounding GenAI have been associated with reduced self-investment \cite{zhao_employees_2024}.

\subsubsection{Task and Role Reconfigurations: Behavior and Adjustments} \label{subsec:response-behavior}

OWs who were already employed in their bridge roles responded to concerns surrounding GenAI by reconfiguring their practices through boundary work, which involves the deliberate construction and maintenance of distinctions that define where one's expertise begins and ends, and where other actors or tools may operate \cite{Gieryn_1999, Farchi2023}. 


Broadly, we observed two distinct approaches to boundary work that OWs employed depending on their career trajectory and the nature of their bridge roles. \textit{Flexible boundaries} involved continuous experimentation and recalibration, particularly among OWs who had transitioned to adjacent or different fields from their careers. \textit{Rigid boundaries} involved firm demarcation between expertise domains and delegation zones, particularly among OWs working in bridge roles similar to their career positions. 
In some cases, OWs combined elements of both. 
P13, for instance, materialized boundaries through a design-thinking framework that structured how students could integrate GenAI while preserving independent  skills, reflecting an approach that was firm in its pedagogical principles but flexible in its incorporation of new tools:
\begin{quote}
    \textit{``Comfy-UI and MidJourney are the tools that we were trying with our students \dots you give a prompt idea, and it can generate sketches and translates it to renders. Students can come up with many of these easily \dots but how they build the idea step-by-step instead of directly asking GenAI to give them an idea is important \dots [this is] how they can govern it \dots I want to teach them how to build the ideas by balancing the use of technology and their own thinking \dots design-thinking with GenAI.''}
\end{quote}
In both cases, OWs drew on their accumulated experience to identify where their contributions remained essential and where GenAI could be integrated. 

\paragraph{Flexible Boundaries} \label{subsec:flexible}
Several OWs, especially those who transitioned to an adjacent or a different bridge role compared to their career job, incorporated flexibility in their boundaries, repeatedly evaluating market forces and reconfiguring their practices in response to emerging developments. The steps to determine these boundaries often involved conducting informal experiments by \textit{delegating} tasks to GenAI, testing the resulting outputs, and comparing them with their own expertise to identify where human skills could still provide distinct value. P5, a linguist who started bridge employment as a voice-over artist, described in detail how she benchmarked her narration skills against three GenAI tools and used these comparisons to carve out a niche within a rapidly disrupted field:

\begin{quote}
     \textit{``I did a comparison between Murf, ElevenLabs, and Wellsaid labs. Synthetic voices of Murph turned out to be very bad. ElevenLabs has a very good voice cloning capability \dots I did a baseline study against my voice. But their sense of the pauses, the sense of emphasis and flow is lacking. \dots Clients still want people to create video narration using synthetic voice in these tools because its quick and cheap \dots to fill the quality gap, I am experimenting by taking synthetic voice but then I am massaging it myself and making it more life-like. I am putting in something extra [expertise] where GenAI is not able to do it.''}
\end{quote}

By elaborating on what was missing in synthetic voices and where an expert human voice could add “something extra,” P5 framed flexible boundaries as embodying resilience in GenAI-driven markets. OWs also used GenAI as a delegation tool to compare outputs from different GenAI systems (e.g., ChatGPT, Perplexity, Grok) to cross-validate results, extract diverse perspectives, or identify blind spots. They believed this comparison strategy offered them greater control over the final content using this process while offsetting the limits of a single system. P20, an entrepreneur who turned consultant as part of their bridge work, explained:
\begin{quote}
    \textit{``I find that it [single system] is not enough. If I put a word like 'procurement' \dots ChatGPT might give me eight points, Perplexity might give me nine points or seven points. Grok giving me 5 points. \dots Sometimes I find there are differences \dots if I notice and take advantage it can help me give an edge.''}
\end{quote}

Other OWs also implemented flexible boundaries, particularly in unstructured tasks where they could not define a clear scope in advance. Such tasks were ill-defined, variable, and context-dependent \cite{suchman2007human}. Examples included managing client expectations, risk assessment and management, and administrative decisions. Participants described how they thought of ways to carve out spaces within these unstructured tasks that allowed them to capitalize on their experience and tacit knowledge, while using GenAI as a \textit{supporting} rather than delegation tool. For instance, P17 pointed to the changing nature of scrum management and GenAI integration in her bridge work as an example of how she adopted GenAI support in a flexible manner:

\begin{quote}
    \textit{``For scrum, AI can actually do end-to-end life cycle management better. It is just a regular process. I will be using my expertise to focus on the product design itself \dots But some things [in scrum lifecycle] need more day-to-day understanding of the system to take a decision at a critical moment. Depending on the criticality, I have to reallocate your [resources] without notice. I have to re-scope something and then do it. I can do it better with AI.”}
\end{quote}

P17’s boundary work reflected a flexible approach that shifted according to how unstructured the task was and how she assessed GenAI’s capabilities in supporting it. 

Although it is common for OWs to make adjustments to their bridge jobs, we found on several occasions (n=5), OWs took additional risk by employing flexible boundaries to reduce their hesitancy in learning and applying their knowledge to new domains. They demonstrated curiosity by combining GenAI’s theoretical knowledge with their own tacit expertise. This combination enabled them to navigate periods of indeterminacy by treating GenAI as a \textit{collaborator} to explore unfamiliar domain knowledge while relying on their experience to contextualize, evaluate, and verify it, thereby increasing their confidence in entering new domains and markets. For example, P15 a lawyer who started working for an American freelance legal platform, elaborated:
\begin{quote}
    \textit{``I am a startup advisor \dots when a US client is approaching me for some work [saying], `Can you do this notice of motion \dots because I have filed a Q64 form.' I will not know what American forms [are] \dots [clients] might have queries, where I don't have awareness. The first thing that I would do is I will put that question to ChatGPT and I would ask \dots ``Please explain this to me. What does the client need?'' Once I understand the question, I can guide the client better.''}
\end{quote}

P15 shared how using ChatGPT enabled her to compete in new markets despite having limited domain knowledge, by combining GenAI content expertise with the vast experience she accumulated while engaging with Indian companies in her career. 
Keeping flexible boundaries opened new possibilities, such as expanding market reach or trying unfamiliar projects, which was enabled, to an extent, by adapting GenAI.

Across these examples, flexible boundary work involved OWs modulating the role they assigned to GenAI along a spectrum, from delegation (assigning well-understood tasks and benchmarking outputs), to support (using GenAI to handle routine components of complex tasks), to collaboration (drawing on GenAI to access unfamiliar knowledge domains). 
They retained their evaluative authority throughout by using their accumulated experience as the benchmark against which those outputs were assessed.

\paragraph{Rigid Boundaries}

Other OWs, especially participants working in similar bridge roles as their career jobs, enacted rigid boundaries between themselves and GenAI. They firmly believed that certain tasks were rooted in the expertise they developed and the tacit knowledge they had accumulated throughout their career, regardless of the level of GenAI integration. They demarcated these tasks from those that were declarative and procedural, with well-defined scope and clear outcomes. These included budgeting, copyediting, work communication, and reporting. OWs considered such tasks an inefficient use of their professional experience, since they did not draw on their accumulated expertise. They saw such tasks fit to delegate to AI. For instance, P18, who started his own financial consultancy as part of his bridge employment, explained GenAI's use case in well-scoped tasks:
\begin{quote}
    \textit{``In the next five years, AI will be able to [do] budgeting, financial reporting, forecasting, because they are about specific parameters and actuals. \dots Once ChatGPT will start looking at it can easily identify patterns and replicate it. It is learning from me. These tasks don't require [my] emotional intelligence or [my] experience on how to react \dots its required in developing financial strategy.''}
\end{quote}

Another category of tasks around which OWs drew rigid boundaries was decision-making. They regarded these tasks as grounded in their accumulated implicit knowledge and professional judgment, making them central to their occupational identity. Consequently, OWs envisioned that decision-making would remain under their control while using GenAI to handle explicit tasks that made the decision process easier. For example, P18, when discussing financial strategy, explained how GenAI could serve as an enabler by supporting critical choices in selecting a strategy:
\begin{quote}
    \textit{“AI will be my enabler. But it cannot make decisions on my behalf, ultimately, it is me who has to make that decision \dots AI will do only what it has been fed. So, it will go to the entire internet, look at the trends, probabilities, and what people have said, and come out with an answer, it is not going to create something of its own \dots it is good in getting you all the research or summarizing those things. But if you expect that it will give you some kind of a blue ocean strategy, no, it will not \dots thats my expertise.”}
\end{quote}

This form of tool reconfiguration also enabled OWs to apply their experiential knowledge to a large volume of structured tasks.

OWs felt that such adjustments to their work to perform a large volume of tasks allowed them to compete with the younger workforce, who were using GenAI to increase their productivity and capitalize on new opportunities. P3, described her process of learning relevant terminologies to perform a systematic review of a large volume of papers using ChatGPT:

\begin{quote}
    \textit{``They asked me to utilize ChatGPT to evaluate the inclusion and exclusion criteria [for literature review]. They wanted 100 papers to be screened in one hour. That was too much, actually \dots around 40,000 papers had to be completed within 15 days. It is easy for me because I used ChatGPT and learnt from it. Whatever questions I see \dots I simply cut and paste the words and then ask what this means, whether it is related to Quantitative study, whether it is about SMEs [Small and Medium Enterprises].''}
\end{quote}

Interestingly, certain OWs felt it was easier to manage such rigid boundaries with GenAI as opposed to other employees. We also saw a few OWs taking strong decisions to eliminate roles and delegate tasks to GenAI. Especially in senior or decision-making roles, OWs described choosing not to hire workers for tasks they deemed easily accomplished by automation. P19, for example, described offloading content marketing to GenAI tools since marketing was not a high priority in their organization:

\begin{quote}
    \textit{``\dots for promotional things \dots you [are] doing voice recording, you [are] putting that on the platforms, you [are] sending it to LinkedIn, you [are] putting on social media, it is so much work you will have to do, unnecessarily. You may have to even hire some people to do that job. Currently, I don't need such resources at all. And these [AI] tools are enough for me to do that work. Because people are going to read it only for one minute. We don't need to put [the] effort of so many people for that.''}
\end{quote}

Through rigid boundaries, OWs established clear domains where their expertise remained central while strategically delegating routine tasks to GenAI. 
This approach enabled them to maintain professional identity and decision-making authority while leveraging GenAI to increase productivity on structured tasks. 
However, as P19's example illustrates, rigid boundaries also had implications for younger workers, as some OWs in managerial positions chose to automate tasks rather than hire employees, a pattern that raises questions about the broader labor market effects of GenAI adoption on bridge workers looking for new opportunities.
\section{Discussion}
Existing scholarship on the future of work in the age of GenAI has largely examined two parallel concerns around the economic outcomes of GenAI adoption, such as gains or losses in productivity \cite{brynjolfsson_productivity_2021}, and the reorganization of workplace processes and roles \cite{frank2019toward}. 
While these studies illuminate top-down transformations, the impact on invisible human labor to adapt to GenAI, particularly for vulnerable communities such as older workers \cite{brewer_why_2016}, remains underexplored. Our study centers on these lived experiences by examining how OWs navigate GenAI-driven disruptions in their pursuit of bridge employment.

Our findings reveal concerns OWs experienced as GenAI was integrated into their bridge employment roles, alongside their agentic responses by reconfiguring their work practices to address these concerns. 
In this section, we interpret these findings to (1) demonstrate how our work extends bridge employment theory, (2) formalize the notion of AI resilience in the context of OWs and the future of work, showing that many resilience responses occurred in isolation at the micro-level, increasing the risk of burnout, and (3) offer recommendations for distributing this adaptive burden across collective (meso) and institutional (macro) levels.

\subsection{Extending Bridge Employment Theory in the Context of GenAI} 

Continuity theory of aging suggests that OWs transitioning from career jobs to permanent retirement seek ways to sustain the psychosocial and behavioral patterns that provided stability and meaning during their careers \cite{Atchley_1989}. Bridge employment offers one such pathway, with OWs seeking \textit{temporal stability} through predictable work trajectories and \textit{structural stability} through work arrangements that allow them to leverage their skills with reduced intensity and greater flexibility.

Our findings show that GenAI disrupted both these forms of stability. At the intention stage, GenAI made transition pathways into bridge roles unpredictable (Section \ref{subsec:concern-planning}). For OWs already engaged in bridge employment, GenAI further destabilized temporal expectations by making the duration of bridge roles uncertain during the behavior and adjustment stage, increasing exposure to competition and vulnerability to replacement (Section \ref{subsec:concern-behavior}).
These findings add complexity to previously identified temporal challenges in the bridge employment and HCI literature that destabilize both the planning stage (e.g., age of entry into bridge employment \cite{Maestas_2010}, time required to learn new technologies \cite{Guo_2017}) and the behavior stage (e.g., health issues \cite{Zhan2009}, technostress \cite{ravsticova2025}).

GenAI also undermined structural stability during the planning phase by decoupling and devaluing skills, such as tacit expertise, that OWs perceived as central to their target bridge roles (Section \ref{subsec:concern-planning}). 
In doing so, GenAI blurred role boundaries that previously enabled OWs to distinguish themselves from a younger workforce during the behavior and adjustment stage, pushing them into direct competition for opportunities, resources, and outcomes (Section \ref{subsec:concern-behavior}). 

GenAI differed from prior technological disruptions because of the simultaneity of its effects across the decision-making process. 
Prior technologies tended to disrupt specific stages in relative isolation \cite{beehr2015working}. 
For example, online job search platforms affected planning \cite{Broder2025}, workplace software affected behavior \cite{Guo_2017}, and algorithmic management impacted adjustment \cite{Segkouli2023}. 
GenAI's general-purpose nature means that it simultaneously reshapes what skills matter (planning), destabilizes role trajectories (intention), devalues tacit knowledge in real-time (behavior), and demands continuous recalibration (adjustment). 
This simultaneity transforms bridge employment decision-making from a staged transition into the recursive, cyclical process we observed.

These disruptions operate within a broader context of structural vulnerability. 
As discussed in Section~\ref{sec:lit_india}, the erosion of institutional retirement protections in India means that many OWs lack the financial security to simply withdraw from bridge employment when conditions deteriorate, compounding the instability that GenAI introduces.

A direct consequence of this uniform disruption is the breakdown of previously linear decision pathways, giving rise to more cyclical processes. 
In earlier models of bridge employment, OWs rarely revisited their intentions once decisions were formed \cite{Beehr2000}. 
Similarly, after securing bridge employment, adjustments were typically periodic and limited in scope \cite{beehr2015working}. 
However, with GenAI, we observed that these stages became recursive. 
OWs continuously recalibrated their intentions as planning evolved, while evolving intentions simultaneously informed planning. 
Likewise, after entering bridge employment, OWs engaged in ongoing adjustments that reshaped their work behaviors. 
These findings demonstrate a shift in bridge employment decision-making toward a more dynamic and iterative process than described in prior literature.
\begin{figure*}
    \centering
    \includegraphics[width=.7\linewidth]{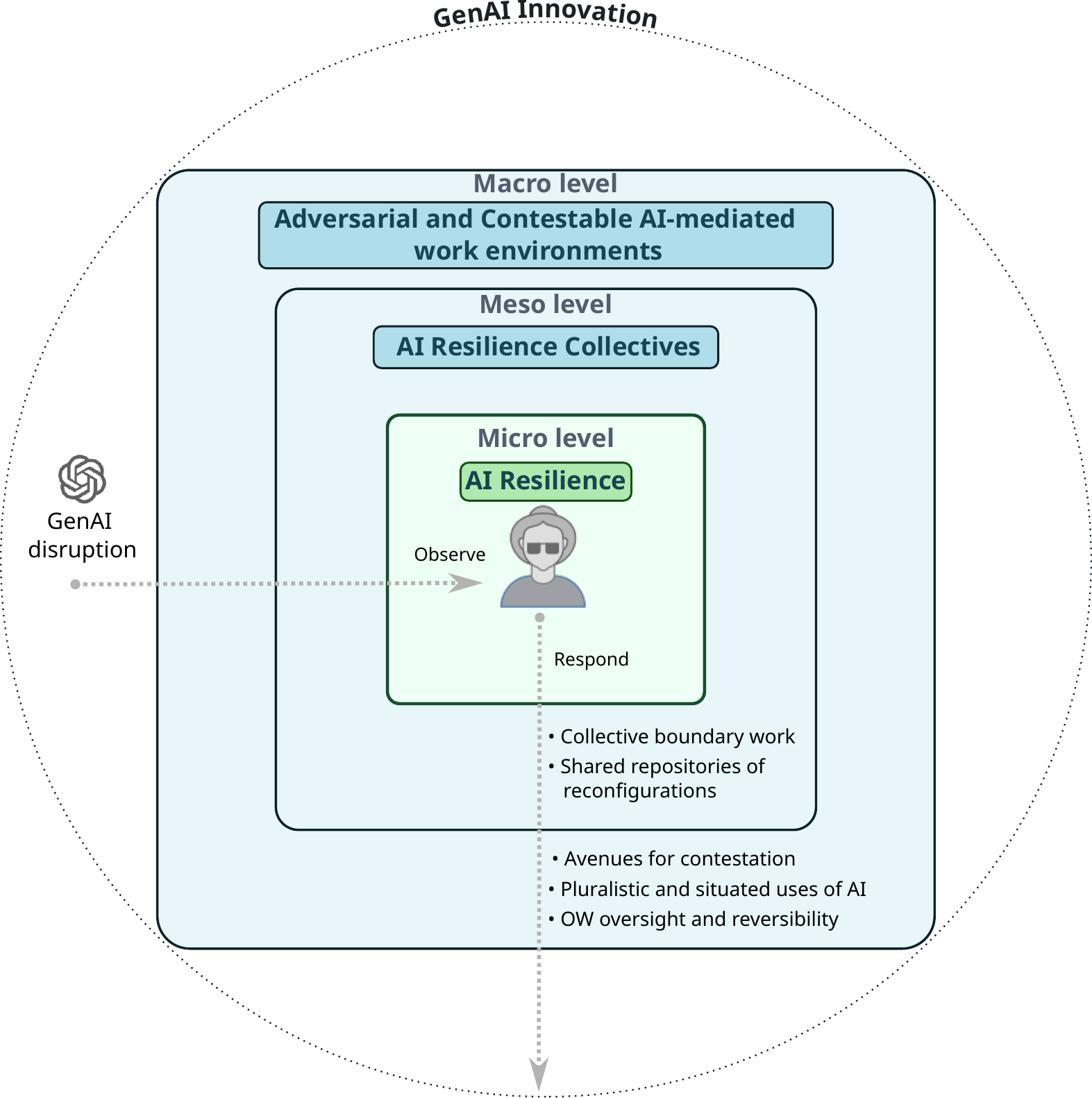}
    \caption{Figure depicting micro-, meso-, and macro-level practices in bridge employment responses to GenAI disruption. The micro-level (innermost layer, green) illustrates AI resilience practices observed in our findings. The meso-level (second layer, blue) represents collective responses, while the macro-level (outer layer, blue) reflects adversarial and contestable initiatives that should be integrated into AI-mediated work environments.}
    \label{fig:new-model}
\end{figure*}
\subsection{Formalizing OWs' AI-Resilience Pathways: A Micro-Level Response}
OWs responded to GenAI disturbances by proactively reconfiguring their practices and reorienting their work identities as a way to bring stability back into their bridge employment. 
Such coping practices are central to \textit{technological resilience}, defined in  the information studies literature as an individual's ability to cope with and thrive in response to external sociotechnical shocks \cite{Heeks2019}. 
This is also in line with how resilience is seen in the gerontology literature as older individuals' ability to overcome adversity through proactive coping ~\cite{van2013ability}. In this context, resilience manifests when OWs recognize and embrace vulnerabilities introduced by shocks and actively work to restore stability and continuity in their work lives ~\cite{Wiles2012}.

We draw on both of these scholarships to formalize the mechanisms of AI resilience among older workers through two key processes. 
The first process was OWs \textit{observing} how GenAI was impacting different aspects of their bridge employment decision-making (innermost layer in Figure \ref{fig:new-model}). 
It involved (1) recognizing how GenAI was being broadly deployed in the bridge employment they were pursuing and (2) contextualizing the implications of that deployment for their own bridge work, including potential benefits and harms. 
For example, in our study, OWs observed that the pace of GenAI integration was far more unsettling than previous waves of technology implementation (Section \ref{subsec:concern-planning}). 
At the same time, they evaluated the impact of such differences on their own jobs, such as increased GenAI training workshops in their own bridge role (Section \ref{subsec:concern-planning}). 
It is also within this process that OWs acknowledged their resultant vulnerabilities and embraced them as part of their responses.

These actions of recognizing and contextualizing parallel the \textit{notice} and \textit{judge} steps that ~\citet{glassman2024ai} argue AI-resilient interfaces should support. However, where Glassman et al.'s framework operates at the interface level---concerned with how specific systems can be designed to support user awareness---the observing process we document operates at the level of professional practice. OWs were not merely noticing how a particular tool behaved; they were assessing how GenAI was restructuring their entire occupational landscape, including skill valuations, competitive dynamics, and role trajectories. This broader scope of observation is what enabled the strategic responses that followed.

The second process was OWs \textit{responding} to anticipated and current vulnerabilities by developing adaptive strategies that either incorporated or resisted GenAI as part of shaping their bridge employment. These responses took structurally different forms depending on career trajectory. OWs who transitioned to adjacent or new fields enacted flexible boundaries, continuously experimenting with and recalibrating GenAI's role---using it variously as a delegation tool, a support mechanism, or a collaborator (Section~\ref{subsec:flexible}). OWs who remained in roles similar to their career positions enacted rigid boundaries, firmly demarcating domains of human expertise from tasks delegable to GenAI. In both cases, OWs drew on accumulated tacit knowledge as the evaluative benchmark against which GenAI outputs were assessed, consistent with resilience scholarship on how older individuals navigate unanticipated life changes ~\cite{Magis_2010, van2013ability}.

Our findings also speak to emerging concerns about de-skilling and cognitive offloading in GenAI-mediated work ~\cite{shukla_-skilling_2025, Lee_2025}. Where studies of younger knowledge workers have documented decreased critical thinking and increased dependence on AI outputs, the OWs in our study exhibited the opposite pattern: they maintained evaluative authority over GenAI, using it instrumentally while preserving their own judgment as the primary basis for professional decisions. This contrast suggests that accumulated professional experience may function as a buffer against the de-skilling effects documented in other populations, a possibility that warrants further investigation across different age groups and career stages.

Framing OWs' responses as AI resilience rather than technology adoption or resistance offers distinct analytical advantages. Adoption frameworks treat the endpoint as integration or rejection of a technology ~\cite{niehaves_internet_2014}. Resistance frameworks foreground opposition. Neither captures the iterative, ongoing negotiation we observed, in which OWs simultaneously incorporated GenAI into some tasks, resisted it in others, and continuously revised these boundaries as conditions changed. AI resilience, by contrast, treats the relationship between workers and GenAI as fundamentally dynamic: not a decision to be made once but a practice to be sustained over time. This framing also centers the worker's accumulated resources---professional judgment, tacit knowledge, experiential breadth---as active assets in navigating disruption, rather than treating age-related experience as a deficit to be overcome through training or adaptation.

\subsection{Promoting Collective Resilience among OWs: A Meso-Level Response}

While the AI resilience practices we observed operated at the individual level, our findings point to the need for complementary mechanisms at higher levels of organization. OWs formulated their strategies by drawing on personal strengths and capacities rather than social or organizational support. 
These findings diverge from prior studies of OW responses to technological disruption, such as computers ~\cite{Lee2008}, where steep learning curves constrained OWs' ability to formulate comparable micro-level adaptations. 
In contrast, GenAI's lower learning curve allowed OWs to respond earlier and through incremental micro-level adjustments embedded in ongoing work. 
However, relying solely on individual adaptation in the face of persistent disruption can impose disproportionate cognitive and emotional burdens on OWs, particularly under conditions of rapid technological change ~\cite{Stanskova2025}.

A complementary approach is to establish bottom-up, meso-level mechanisms that leverage the collective strengths of OWs while bolstering micro-level resilience (middle layer in Figure \ref{fig:new-model}). 
Communities of practice provide a useful lens for understanding how such collective mechanisms may emerge in OW contexts ~\cite{wenger_communities_1998, brown_organizational_1991, millen_improving_2003}. 
Prior research suggests that successful communities of practice among older individuals rely on social networks to support identity continuity through shared negotiation and mutual recognition ~\cite{Nimrod_2014}. These communities can enable OWs to collectively discuss, renegotiate, and reframe their professional identities and skills in response to GenAI-driven disruption, rather than requiring individuals to continually justify their professional identity and the value of their skills in isolation.

Such communities can also support collective boundary work ~\cite{Contu2014}, enabling OW work communities to jointly articulate and legitimize modes of participation that distinguish OW expertise from GenAI capabilities. 
In addition, shared repertoires, including routines, narratives, heuristics, and artifacts generated as part of community interactions, play a critical role in sustaining professional competence and identity over time. 
In the context of GenAI-driven skill reconfiguration, such shared repertoires can help preserve tacit and experiential knowledge as collective assets by making domain-specific judgment and practices visible and socially validated.

\subsection{Promoting Adversarial and Contestable AI-Mediated Work Ecosystems for OWs: A Macro-Level Response}
OW resilience strategies were also distinctly cautious and adversarial, allowing them to resist dominant efficiency-driven framings of GenAI and instead build on assets such as experience and tacit knowledge. This was enabled in part by GenAI's nature as a general-purpose, personalizable technology, which supports proactive adversarial crafting ~\cite{Var2026}.

We use the term \textit{adversarial}, building on ~\citet{disalvo_adversarial_2015}'s work, to describe a mode of engagement in which workers actively contest dominant framings of AI, set boundaries around its use, and reassert human judgment as a central organizing principle. 
The OWs in our study approached GenAI from this adversarial stance, shaped by their deep experiential knowledge and long-standing professional identities that enabled them to surface concerns and identify directions for action. 
Adversariality, in this sense, is a design principle for ensuring dignity and inclusion in the future of work. 
The boundary work practices documented in our findings, particularly the rigid boundaries through which OWs reasserted decision-making authority over GenAI, suggest that adversarial engagement has to go beyond individual disposition and into the structural level. 
We call for AI-mediated workplace structures at the macro level that explicitly support and bolster adversarial and contestable engagement with AI at micro and meso levels (outer layer in Figure~\ref{fig:new-model}).

We envision such macro-level ecosystems promoting adversarial and contestable work environments in three key ways. 
First, they should provide institutionalized avenues for contestation, enabling OWs to act as legitimate peripheral participants ~\cite{lave1991situated} who can question, override, and renegotiate AI-mediated decisions rather than comply by default. 
This aligns with calls for governance mechanisms that support contestable systems ~\cite{lyons2021}. 
Second, such ecosystems should legitimize pluralistic and situated uses of AI by OWs (i.e., as a support, manager, delegation, or collaboration), while recognizing that appropriate forms of AI integration vary across roles, bridge employment stages, and worker values, consistent with agonistic perspectives in HCI ~\cite{disalvo_adversarial_2015}. 
These systems can also act as an avenue to legitimize such practices through formal programs. 
Third, they should embed human oversight and reversibility as structural expectations, ensuring that OWs experience a sense of control over AI systems rather than encountering them as fixed authorities, a concern emphasized in macro-level critiques of misaligned AI ecosystems ~\cite{Kane2021}. 
These qualities can create work environments in which OWs' resilience practices are not exceptional acts of individual resistance, but supported modes of participation that enable continued stability, dignity, and thriving in AI-mediated work.
\subsection{Limitations and Future Work}
Our study has several limitations. While it includes a diverse cross-section of OWs engaged in knowledge and creative work across different stages of the bridge employment process, the insights are shaped by the methodological constraints typical of qualitative research, including a relatively small sample size. 
Our decision to study OWs in India was motivated by their high exposure to GenAI and the global relevance and generalizability of their professional roles. While this study does not center on sociocultural dynamics specific to the Indian context, these dynamics represent an important direction for future research. Additional research is also needed to quantitatively examine the impact of GenAI on the bridge employment decision-making processes, to track how boundary work evolves longitudinally. 
Lastly, more research is required to understand how OWs envision their work futures in this age of AI-mediated work.

\section{Conclusion}
In this study, we examined the concerns and responses of older workers (OWs)  experiencing the rapid proliferation of GenAI as they sought or engaged in bridge employment. We identified distinct temporal and structural disruptions that OWs encountered throughout the bridge employment decision-making process. These disruptions contributed to fears and vulnerabilities that undermined the stability and continuity that bridge employment would otherwise provide prior to full retirement. Moreover, GenAI’s persistent disruptions transformed bridge employment decision-making into an ongoing process of negotiation and adaptation, diverging from established models that frame this transition as largely linear.

Despite these challenges, OWs responded by reorienting their professional roles and reconfiguring their work practices in ways that helped restore stability. By examining these bottom-up strategies, our study formalizes the notion of AI resilience in the context of OWs and the future of work. Within this framework, we identify two key processes through which OWs navigate sociotechnical disruption: \textit{observing} and \textit{responding}. 
Through these processes, OWs assess emerging risks and opportunities associated 
with GenAI and develop adaptive strategies, enacted through flexible and rigid 
boundary work.

At the same time, because many OWs enacted these AI resilience practices independently at the micro-level, they faced heightened risks of cumulative cognitive and emotional strain. Drawing on a micro-, meso-, and macro-level framework of bridge employment, we conclude by outlining pathways for distributing this burden more broadly. These include cultivating collective resilience infrastructures at the meso-level through Communities of Practice, as well as promoting AI-mediated work environments at the macro-level that support adversarial and contestable forms of engagement for OWs.

\section{AI Disclosure Statement}
We are grateful to the participants for sharing their thoughts and giving us their time. We also acknowledge the use of GenAI as a copyediting tool to improve spelling, grammar, punctuation, clarity, and engagement.


\bibliographystyle{ACM-Reference-Format}
\bibliography{mainbib}

@article{Topa_2014, title={Bridge employment quality and its impact on retirement adjustment: A structural equation model with SHARE panel data}, volume={35}, ISSN={0143-831X, 1461-7099}, url={https://journals.sagepub.com/doi/10.1177/0143831X12475242}, DOI={10.1177/0143831X12475242},number={2}, journal={Economic and Industrial Democracy}, author={Topa, Gabriela and Alcover, Carlos-María and Moriano, Juan A and Depolo, Marco}, year={2014}, month=may, pages={225–244}, language={en} }

@article{Dingemans_Henkens_2014, title={Involuntary retirement, bridge employment, and satisfaction with life: A longitudinal investigation: Retirement Transitions and Life Satisfaction}, volume={35}, ISSN={08943796}, url={https://onlinelibrary.wiley.com/doi/10.1002/job.1914}, DOI={10.1002/job.1914}, number={4}, journal={Journal of Organizational Behavior}, author={Dingemans, Ellen and Henkens, Kène}, year={2014}, month=may, pages={575–591}, language={en} }

@article{Drentea_2002, title={Retirement and Mental Health}, volume={14}, rights={https://journals.sagepub.com/page/policies/text-and-data-mining-license}, ISSN={0898-2643, 1552-6887}, url={https://journals.sagepub.com/doi/10.1177/089826430201400201}, DOI={10.1177/089826430201400201},  number={2}, journal={Journal of Aging and Health}, author={Drentea, Patricia}, year={2002}, month=may, pages={167–194}, language={en} }

@article{Nakai_2011, title={Profiles of mature job seekers: Connecting needs and desires to work characteristics}, volume={32}, rights={http://onlinelibrary.wiley.com/termsAndConditions#vor}, ISSN={0894-3796, 1099-1379}, url={https://onlinelibrary.wiley.com/doi/10.1002/job.697}, DOI={10.1002/job.697},  number={2}, journal={Journal of Organizational Behavior}, author={Nakai, Yoshie and Chang, Boin and Snell, Andrea F. and Fluckinger, Chris D.}, year={2011}, month=feb, pages={155–172}, language={en} }

@article{Shultz_2003, title={Work Related Attitudes of Naval Officers before and after Retirement}, volume={57}, rights={https://journals.sagepub.com/page/policies/text-and-data-mining-license}, ISSN={0091-4150, 1541-3535}, url={https://journals.sagepub.com/doi/10.2190/B4RV-PW54-J29B-FNV1}, DOI={10.2190/B4RV-PW54-J29B-FNV1},  number={3}, journal={The International Journal of Aging and Human Development}, author={Shultz, Kenneth S. and Taylor, Mary Anne and Morrison, Robert F.}, year={2003}, month=oct, pages={259–274}, language={en} }

@book{Alcover_2014,
  title={Bridge employment: A research handbook},
  author={Alcover, Carlos-Maria and Topa, Gabriela and Parry, Emma and Fraccaroli, Franco and Depolo, Marco},
  year={2014},
  publisher={Routledge}
}

@article{doeringer1990business,
  title={Business necessity, bridge jobs, and the nonbureaucratic firm},
  author={Doeringer, Peter B and Terkla, David G},
  journal={Bridges to retirement: Older workers in a changing labor market},
  pages={146--171},
  year={1990},
  publisher={Cornell University, ILR Press Ithaca, NY}
}

@article{Kantarci2008, title={Gradual Retirement: Preferences and Limitations}, volume={156}, ISSN={0013-063X, 1572-9982}, url={http://link.springer.com/10.1007/s10645-008-9086-1}, DOI={10.1007/s10645-008-9086-1}, number={2}, journal={De Economist}, author={Kantarci, Tunga and Van Soest, Arthur}, year={2008}, month=jun, pages={113–144}, language={en} }

@online{EEOC1967,
  author       = {{U.S. Equal Employment Opportunity Commission}},
  title        = {Age Discrimination in Employment Act of 1967},
  year         = {2024},
  url          = {https://www.eeoc.gov/statutes/age-discrimination-employment-act-1967},
  urldate      = {2026-03-30},
  note         = {Accessed 30 March 2026}
}

@article{Trier2012, title={Framing the telephone interview as a participant-centred tool for qualitative research: a methodological discussion}, volume={12}, ISSN={1468-7941, 1741-3109}, url={https://journals.sagepub.com/doi/10.1177/1468794112439005}, DOI={10.1177/1468794112439005}, number={6}, journal={Qualitative Research}, author={Trier-Bieniek, Adrienne}, year={2012}, month=dec, pages={630–644}, language={en} }

@article{Creswell2000, title={Determining Validity in Qualitative Inquiry}, volume={39}, ISSN={0040-5841, 1543-0421}, url={http://www.tandfonline.com/doi/abs/10.1207/s15430421tip3903_2}, DOI={10.1207/s15430421tip3903_2}, number={3}, journal={Theory Into Practice}, author={Creswell, John W. and Miller, Dana L.}, year={2000}, month=aug, pages={124–130}, language={en} }

@article{Soden2024, title={Evaluating Interpretive Research in HCI}, volume={31}, ISSN={1072-5520, 1558-3449}, url={https://dl.acm.org/doi/10.1145/3633200}, DOI={10.1145/3633200}, number={1}, journal={Interactions}, author={Soden, Robert and Toombs, Austin and Thomas, Michaelanne}, year={2024}, month=jan, pages={38–42}, language={en} }

@article{smith2024qual,
  title={Qualitative psychology: A practical guide to research methods},
  author={Smith, Jonathan A},
  year={2024},
  publisher={Sage Publications Ltd}
}

@article{shultz2011psychological,
  title={Psychological perspectives on the changing nature of retirement.},
  author={Shultz, Kenneth S and Wang, Mo},
  journal={American psychologist},
  volume={66},
  number={3},
  pages={170},
  year={2011},
  publisher={American Psychological Association}
}

@book{hedge2012oxford,
  title={The Oxford handbook of work and aging},
  author={Hedge, Jerry W and Borman, Walter C},
  year={2012},
  publisher={Oxford University Press, USA}
}

@article{gobeski2009retirees,
  title={How retirees work: Predictors of different types of bridge employment},
  author={Gobeski, Kirsten T and Beehr, Terry A},
  journal={Journal of Organizational Behavior: The International Journal of Industrial, Occupational and Organizational Psychology and Behavior},
  volume={30},
  number={3},
  pages={401--425},
  year={2009},
  publisher={Wiley Online Library}
}

@incollection{henretta2018life,
  title={The life-course perspective on work and retirement},
  author={Henretta, John C},
  booktitle={Invitation to the life course},
  pages={85--105},
doi = {10.4324/9781315224206},
url = {https://doi.org/10.4324/9781315224206},
  year={2018},
  publisher={Routledge}
}

@article{beehr2015working,
  title={Working after retirement: Features of bridge employment and research directions},
  author={Beehr, Terry A and Bennett, Misty M},
  journal={Work, Aging and Retirement},
  volume={1},
  number={1},
  pages={112--128},
  year={2015},
  publisher={Oxford University Press US}
}

@article{righi2017we,
  title={When we talk about older people in HCI, who are we talking about? Towards a ‘turn to community’in the design of technologies for a growing ageing population},
  author={Righi, Valeria and Sayago, Sergio and Blat, Josep},
  journal={International Journal of Human-Computer Studies},
  volume={108},
  pages={15--31},
  year={2017},
  publisher={Elsevier}
}

@misc{glassman2024ai,
      title={AI-Resilient Interfaces}, 
      author={Elena L. Glassman and Ziwei Gu and Jonathan K. Kummerfeld},
      year={2024},
      eprint={2405.08447},
      archivePrefix={arXiv},
      primaryClass={cs.HC},
      url={https://arxiv.org/abs/2405.08447}, 
}

@article{Heeks2019, title={Conceptualising the link between information systems and resilience: A developing country field study}, volume={29}, rights={http://onlinelibrary.wiley.com/termsAndConditions#vor}, ISSN={1350-1917, 1365-2575}, url={https://onlinelibrary.wiley.com/doi/10.1111/isj.12177}, DOI={10.1111/isj.12177}, number={1}, journal={Information Systems Journal}, author={Heeks, Richard and Ospina, Angelica V.}, year={2019}, month=jan, pages={70–96}, language={en} }

@article{Magis_2010, title={Community Resilience: An Indicator of Social Sustainability}, volume={23}, ISSN={0894-1920, 1521-0723}, url={http://www.tandfonline.com/doi/abs/10.1080/08941920903305674}, DOI={10.1080/08941920903305674}, number={5}, journal={Society \& Natural Resources}, author={Magis, Kristen}, year={2010}, month=apr, pages={401–416}, language={en} }

@article{Wiles2012, title={Resilience from the point of view of older people: ‘There’s still life beyond a funny knee’}, volume={74}, rights={https://www.elsevier.com/tdm/userlicense/1.0/}, ISSN={02779536}, url={https://linkinghub.elsevier.com/retrieve/pii/S0277953611007155}, DOI={10.1016/j.socscimed.2011.11.005}, number={3}, journal={Social Science \& Medicine}, author={Wiles, Janine L. and Wild, Kirsty and Kerse, Ngaire and Allen, Ruth E.S.}, year={2012}, month=feb, pages={416–424}, language={en} }

@inproceedings{Guo_2017, address={Denver Colorado USA}, title={Older Adults Learning Computer Programming: Motivations, Frustrations, and Design Opportunities}, ISBN={9781450346559}, url={https://dl.acm.org/doi/10.1145/3025453.3025945}, DOI={10.1145/3025453.3025945}, booktitle={Proceedings of the 2017 CHI Conference on Human Factors in Computing Systems}, publisher={ACM}, author={Guo, Philip J.}, year={2017}, month=may, pages={7070–7083}, language={en} }

@article{Maestas_2010, title={Back to Work: Expectations and Realizations of Work after Retirement}, volume={45}, ISSN={0022-166X, 1548-8004}, url={http://jhr.uwpress.org/lookup/doi/10.3368/jhr.45.3.718}, DOI={10.3368/jhr.45.3.718}, number={3}, journal={Journal of Human Resources}, author={Maestas, Nicole}, year={2010}, pages={718–748}, language={en} }

@article{Contu2014, title={On boundaries and difference: Communities of practice and power relations in creative work}, volume={45}, ISSN={1350-5076, 1461-7307}, url={https://journals.sagepub.com/doi/10.1177/1350507612471926}, DOI={10.1177/1350507612471926},  number={3}, journal={Management Learning}, author={Contu, Alessia}, year={2014}, month=july, pages={289–316}, language={en} }

@article{Atchley_1989, title={A Continuity Theory of Normal Aging}, volume={29}, ISSN={0016-9013, 1758-5341}, url={https://academic.oup.com/gerontologist/article-lookup/doi/10.1093/geront/29.2.183}, DOI={10.1093/geront/29.2.183}, number={2}, journal={The Gerontologist}, author={Atchley, R. C.}, year={1989}, month=apr, pages={183–190}, language={en} }

@article{Stanskova2025, title={Adapting to Technological Change: Cognitive and Emotional Response of Older Employees in the Industrial Sector}, rights={http://creativecommons.org/licenses/by/4.0}, url={https://www.preprints.org/manuscript/202504.2571/v1}, DOI={10.20944/preprints202504.2571.v1}, author={Stanskova, Darya}, year={2025}, month=apr }

@article{ravsticova2025,
  title={Technostress Among Older Workers: A Central European Perspective},
  author={Ra{\v{s}}ticov{\'a}, Martina and {\v{S}}{\'a}cha, Jakub and Lakom{\`y}, Martin and Mishra, Pawan Kumar},
  journal={Psychology Research and Behavior Management},
  url = {https://doi.org/10.2147/PRBM.S508500},
  DOI = {10.2147/PRBM.S508500},
  pages={1211--1225},
  year={2025},
  publisher={Taylor \& Francis}
}

@article{Zhan2009, title={Bridge employment and retirees’ health: A longitudinal investigation.}, volume={14}, ISSN={1939-1307, 1076-8998}, url={https://doi.apa.org/doi/10.1037/a0015285}, DOI={10.1037/a0015285}, number={4}, journal={Journal of Occupational Health Psychology}, author={Zhan, Yujie and Wang, Mo and Liu, Songqi and Shultz, Kenneth S.}, year={2009}, pages={374–389}, language={en} }

@article{Lee2008, title={Training Older Workers for Technology-Based Employment}, volume={35}, ISSN={0360-1277, 1521-0472}, url={http://www.tandfonline.com/doi/abs/10.1080/03601270802300091}, DOI={10.1080/03601270802300091}, number={1}, journal={Educational Gerontology}, author={Lee, Chin Chin and Czaja, Sara J. and Sharit, Joseph}, year={2008}, month=dec, pages={15–31}, language={en} }

@report{ilo2024,
  title        = {India Employment Report 2024: Youth Employment, Education, and Skills},
  author       = {{International Labour Organization}},
  institution  = {International Labour Organization},
  year         = {2024},
  url          = {https://www.ilo.org/sites/default/files/2024-08/India%20Employment%20-%20web_8%20April.pdf}
}

@inbook{Law2025, address={Cham}, title={Generative AI and Changing Work: Systematic Review of Practitioner-Led Work Transformations Through the Lens of Job Crafting}, volume={15804}, ISBN={9783031928222}, url={https://link.springer.com/10.1007/978-3-031-92823-9_10}, DOI={10.1007/978-3-031-92823-9_10}, booktitle={HCI in Business, Government and Organizations}, publisher={Springer Nature Switzerland}, author={Law, Matthew and Varanasi, Rama Adithya}, editor={Siau, Keng Leng and Nah, Fiona Fui-Hoon}, year={2025}, pages={131–152}, language={en} }

@article{feldman1994decision,
  title={The decision to retire early: A review and conceptualization},
  author={Feldman, Daniel C},
  journal={Academy of management review},
  volume={19},
  number={2},
  pages={285--311},
  year={1994},
  publisher={Academy of Management Briarcliff Manor, NY 10510}
}

@article{kim2000working,
  title={Working in retirement: The antecedents of bridge employment and its consequences for quality of life in retirement},
  author={Kim, Seongsu and Feldman, Daniel C},
  journal={Academy of management Journal},
  volume={43},
  number={6},
  pages={1195--1210},
  year={2000},
  publisher={Academy of Management Briarcliff Manor, NY 10510}
}

@article{kuoppamaki2021designing,
  title={Designing Kitchen Technologies for Ageing in Place: A Video Study of Older Adults' Cooking at Home},
  author={Kuoppam{\"a}ki, Sanna and Tuncer, Sylvaine and Eriksson, Sara and McMillan, Donald},
  journal={Proceedings of the ACM on interactive, mobile, wearable and ubiquitous technologies},
  volume={5},
  number={2},
  pages={1--19},
  year={2021},
  publisher={ACM New York, NY, USA}
}

@inproceedings{rogers2014never,
  title={Never too old: engaging retired people inventing the future with MaKey MaKey},
  author={Rogers, Yvonne and Paay, Jeni and Brereton, Margot and Vaisutis, Kate L and Marsden, Gary and Vetere, Frank},
  booktitle={Proceedings of the SIGCHI Conference on Human Factors in Computing Systems},
  pages={3913--3922},
  year={2014}
}

@inproceedings{vines2012cheque,
  title={Cheque mates: participatory design of digital payments with eighty somethings},
  author={Vines, John and Blythe, Mark and Dunphy, Paul and Vlachokyriakos, Vasillis and Teece, Isaac and Monk, Andrew and Olivier, Patrick},
  booktitle={Proceedings of the SIGCHI Conference on Human Factors in Computing Systems},
  pages={1189--1198},
  year={2012}
}

@inproceedings{vines2012questionable,
  title={Questionable concepts: critique as resource for designing with eighty somethings},
  author={Vines, John and Blythe, Mark and Lindsay, Stephen and Dunphy, Paul and Monk, Andrew and Olivier, Patrick},
  booktitle={Proceedings of the SIGCHI Conference on Human Factors in Computing Systems},
  pages={1169--1178},
  year={2012}
}

@inproceedings{lindsay2012empathy,
  title={Empathy, participatory design and people with dementia},
  author={Lindsay, Stephen and Brittain, Katie and Jackson, Daniel and Ladha, Cassim and Ladha, Karim and Olivier, Patrick},
  booktitle={Proceedings of the SIGCHI conference on Human factors in computing systems},
  pages={521--530},
  year={2012}
}

@article{ekerdt1990defining,
  title={On defining persons as retired},
  author={Ekerdt, David J and DeViney, Stanley},
  journal={Journal of Aging Studies},
  volume={4},
  number={3},
  pages={211--229},
  year={1990},
  publisher={Elsevier}
}

@book{bell_storytelling_2020,
	address = {New York London},
	edition = {Second edition},
	series = {The teaching/learning social justice series},
	title = {Storytelling for social justice: connecting narrative and the arts in antiracist teaching},
	isbn = {978-1-138-29280-2 978-1-315-10104-0},
	shorttitle = {Storytelling for social justice},
	language = {eng},
	publisher = {Routledge, Taylor \& Francis Group},
	author = {Bell, Lee Anne},
	year = {2020},
}

@article{levy_stereotype_2009,
	title = {Stereotype {Embodiment}},
	volume = {18},
	issn = {0963-7214},
	url = {https://www.ncbi.nlm.nih.gov/pmc/articles/PMC2927354/},
	doi = {10.1111/j.1467-8721.2009.01662.x},
	number = {6},
	urldate = {2025-05-08},
	journal = {Current directions in psychological science},
	author = {Levy, Becca},
	month = dec,
	year = {2009},
	pmid = {20802838},
	pmcid = {PMC2927354},
	pages = {332--336},
}

@article{barnard_learning_2013,
	title = {Learning to use new technologies by older adults: {Perceived} difficulties, experimentation behaviour and usability},
	volume = {29},
	issn = {0747-5632},
	shorttitle = {Learning to use new technologies by older adults},
	url = {https://www.sciencedirect.com/science/article/pii/S0747563213000721},
	doi = {10.1016/j.chb.2013.02.006},
	number = {4},
	urldate = {2025-05-08},
	journal = {Computers in Human Behavior},
	author = {Barnard, Yvonne and Bradley, Mike D. and Hodgson, Frances and Lloyd, Ashley D.},
	month = jul,
	year = {2013},
	pages = {1715--1724},
}

@article{fletcher-watson_strategies_2016,
	title = {Strategies for enhancing success in digital tablet use by older adults: {A} pilot study},
	volume = {15},
	issn = {1569-111X, 1569-1101},
	shorttitle = {Strategies for enhancing success in digital tablet use by older adults},
	url = {https://journal.gerontechnology.org/currentIssueContent.aspx?aid=2464},
	doi = {10.4017/gt.2016.15.3.005.00},
	number = {3},
	urldate = {2025-05-08},
	journal = {Gerontechnology},
	author = {Fletcher-Watson, B. and Crompton, C.J. and Hutchison, M. and Lu, H.},
	month = nov,
	year = {2016},
	pages = {162--170},
}

@techreport{MarshMcLennan2018,
  title        = {The Twin Threats of Aging and Automation},
  author       = {{Marsh \& McLennan Companies}},
  year         = {2018},
  institution  = {Marsh \& McLennan Companies, Asia Pacific Risk Center},
  url          = {https://www.marshmclennan.com/web-assets/insights/publications/2018/dec/workforce-of-the-future/The-Twin-Threats-Of-Aging-And-Automation/The-Twin-Threats-Of-Aging-And-Automation.pdf},
  note         = {Accessed on August 10, 2025}
}

@article{Pak2017, title={The effect of individual differences in working memory in older adults on performance with different degrees of automated technology}, volume={60}, ISSN={0014-0139, 1366-5847}, url={https://www.tandfonline.com/doi/full/10.1080/00140139.2016.1189599}, DOI={10.1080/00140139.2016.1189599}, number={4}, journal={Ergonomics}, author={Pak, Richard and McLaughlin, Anne Collins and Leidheiser, William and Rovira, Ericka}, year={2017}, month=apr, pages={518–532}, language={en} }

@article{Atkinson2016, title={An exploration of older worker flexible working arrangements in smaller firms}, volume={26}, rights={http://creativecommons.org/licenses/by-nc/4.0/}, ISSN={0954-5395, 1748-8583}, url={https://onlinelibrary.wiley.com/doi/10.1111/1748-8583.12074}, DOI={10.1111/1748-8583.12074}, 
number={1}, journal={Human Resource Management Journal}, author={Atkinson, Carol and Sandiford, Peter}, year={2016}, month=jan, pages={12–28}, language={en} }

@article{coudin_help_2010,
	title = {‘{Help} me! {I}’m old!’ {How} negative aging stereotypes create dependency among older adults},
	volume = {14},
	issn = {1360-7863, 1364-6915},
	url = {https://www.tandfonline.com/doi/full/10.1080/13607861003713182},
	doi = {10.1080/13607861003713182},
	language = {en},
	number = {5},
	urldate = {2025-05-08},
	journal = {Aging \& Mental Health},
	author = {Coudin, Geneviève and Alexopoulos, Theodore},
	month = jul,
	year = {2010},
	pages = {516--523},
}

@article{levy_mind_2003,
	title = {Mind {Matters}: {Cognitive} and {Physical} {Effects} of {Aging} {Self}-{Stereotypes}},
	volume = {58},
	issn = {1758-5368, 1079-5014},
	shorttitle = {Mind {Matters}},
	url = {https://academic.oup.com/psychsocgerontology/article-lookup/doi/10.1093/geronb/58.4.P203},
	doi = {10.1093/geronb/58.4.P203},
	language = {en},
	number = {4},
	urldate = {2025-05-08},
	journal = {The Journals of Gerontology: Series B},
	author = {Levy, Becca R.},
	month = jul,
	year = {2003},
	pages = {P203--P211},
}

@article{disalvo_design_2009,
	title = {Design and the {Construction} of {Publics}},
	volume = {25},
	issn = {0747-9360, 1531-4790},
	url = {https://direct.mit.edu/desi/article/25/1/48-63/68938},
	doi = {10.1162/desi.2009.25.1.48},
	language = {en},
	number = {1},
	urldate = {2025-05-08},
	journal = {Design Issues},
	author = {DiSalvo, Carl},
	month = jan,
	year = {2009},
	pages = {48--63},
}

@article{lyons2021,
author = {Lyons, Henrietta and Velloso, Eduardo and Miller, Tim},
title = {Conceptualising Contestability: Perspectives on Contesting Algorithmic Decisions},
year = {2021},
issue_date = {April 2021},
publisher = {Association for Computing Machinery},
address = {New York, NY, USA},
volume = {5},
number = {CSCW1},
url = {https://doi.org/10.1145/3449180},
doi = {10.1145/3449180},
journal = {Proc. ACM Hum.-Comput. Interact.},
month = apr,
articleno = {106},
numpages = {25}
}

@article{Kane2021, title={Avoiding an Oppressive Future of Machine Learning: A Design Theory for Emancipatory assistants}, volume={45}, ISSN={0276-7783, 2162-9730}, url={https://misq.umn.edu/misq/article/45/1/371/1830/Avoiding-an-Oppressive-Future-of-Machine-Learning}, DOI={10.25300/MISQ/2021/1578},number={1}, journal={MIS Quarterly}, author={Kane, Gerald C. and Young, Amber G. and Majchrzak, Ann and Ransbotham, Sam}, year={2021}, month=mar, pages={371–396}, language={en} }

@article{Broder2025, title={Too old to find employment? A novel approach to leverage the power of digital peer groups for older unemployed}, volume={23}, ISSN={1617-9846, 1617-9854}, url={https://link.springer.com/10.1007/s10257-025-00708-3}, DOI={10.1007/s10257-025-00708-3},  number={4}, journal={Information Systems and e-Business Management}, author={Broder, Hanna Rebecca and Förster, Maximilian and Klier, Julia and Klier, Mathias and Sigler, Irina}, year={2025}, month=dec, pages={999–1038}, language={en} }

@book{lave1991situated,
  title={Situated learning: Legitimate peripheral participation},
  author={Lave, Jean and Wenger, Etienne},
  year={1991},
  publisher={Cambridge university press}
}

@article{Segkouli2023, title={Smart Workplaces for older adults: coping ‘ethically’ with technology pervasiveness}, volume={22}, ISSN={1615-5289, 1615-5297}, url={https://link.springer.com/10.1007/s10209-021-00829-9}, DOI={10.1007/s10209-021-00829-9}, number={1}, journal={Universal Access in the Information Society}, author={Segkouli, Sofia and Giakoumis, Dimitrios and Votis, Konstantinos and Triantafyllidis, Andreas and Paliokas, Ioannis and Tzovaras, Dimitrios}, year={2023}, month=mar, pages={37–49}, language={en} }

@article{Nimrod_2014, title={The benefits of and constraints to participation in seniors’ online communities}, volume={33}, ISSN={0261-4367, 1466-4496}, url={http://www.tandfonline.com/doi/abs/10.1080/02614367.2012.697697}, DOI={10.1080/02614367.2012.697697}, number={3}, journal={Leisure Studies}, author={Nimrod, Galit}, year={2014}, month=may, pages={247–266}, language={en} }

@article{Beehr2000, title={Work and Nonwork Predictors of Employees’ Retirement Ages}, volume={57}, rights={https://www.elsevier.com/tdm/userlicense/1.0/}, ISSN={00018791}, url={https://linkinghub.elsevier.com/retrieve/pii/S0001879199917360}, DOI={10.1006/jvbe.1999.1736}, number={2}, journal={Journal of Vocational Behavior}, author={Beehr, Terry A. and Glazer, Sharon and Nielson, Norma L. and Farmer, Suzanne J.}, year={2000}, month=oct, pages={206–225}, language={en} }

@article{Wang2008, title={Antecedents of bridge employment: A longitudinal investigation.}, volume={93}, ISSN={1939-1854, 0021-9010}, url={https://doi.apa.org/doi/10.1037/0021-9010.93.4.818}, DOI={10.1037/0021-9010.93.4.818}, number={4}, journal={Journal of Applied Psychology}, author={Wang, Mo and Zhan, Yujie and Liu, Songqi and Shultz, Kenneth S.}, year={2008}, pages={818–830}, language={en} }

@inproceedings{kobayashi_motivating_2015,
	address = {Vancouver BC Canada},
	title = {Motivating {Multi}-{Generational} {Crowd} {Workers} in {Social}-{Purpose} {Work}},
	isbn = {978-1-4503-2922-4},
	url = {https://dl.acm.org/doi/10.1145/2675133.2675255},
	doi = {10.1145/2675133.2675255},
	language = {en},
	urldate = {2025-05-08},
	booktitle = {Proceedings of the 18th {ACM} {Conference} on {Computer} {Supported} {Cooperative} {Work} \& {Social} {Computing}},
	publisher = {ACM},
	author = {Kobayashi, Masatomo and Arita, Shoma and Itoko, Toshinari and Saito, Shin and Takagi, Hironobu},
	month = feb,
	year = {2015},
	pages = {1813--1824},
}

@article{Ho_Yeung_2021, title={Conflict between younger and older workers: an identity-based approach}, volume={32}, rights={https://www.emerald.com/insight/site-policies}, ISSN={1044-4068, 1044-4068}, url={http://www.emerald.com/ijcma/article/32/1/102-125/120557}, DOI={10.1108/IJCMA-08-2019-0124}, number={1}, journal={International Journal of Conflict Management}, author={Ho, Henry C.Y. and Yeung, Dannii Y.}, year={2021}, month=jan, pages={102–125}, language={en} }

@inproceedings{brewer_why_2016,
	address = {San Jose California USA},
	title = {"{Why} would anybody do this?": {Understanding} {Older} {Adults}' {Motivations} and {Challenges} in {Crowd} {Work}},
	isbn = {978-1-4503-3362-7},
	shorttitle = {"{Why} would anybody do this?},
	url = {https://dl.acm.org/doi/10.1145/2858036.2858198},
	doi = {10.1145/2858036.2858198},
	language = {en},
	urldate = {2025-05-08},
	booktitle = {Proceedings of the 2016 {CHI} {Conference} on {Human} {Factors} in {Computing} {Systems}},
	publisher = {ACM},
	author = {Brewer, Robin and Morris, Meredith Ringel and Piper, Anne Marie},
	month = may,
	year = {2016},
	pages = {2246--2257},
}

@article{niehaves_internet_2014,
	title = {Internet adoption by the elderly: employing {IS} technology acceptance theories for understanding the age-related digital divide},
	volume = {23},
	issn = {1476-9344},
	shorttitle = {Internet adoption by the elderly},
	url = {https://doi.org/10.1057/ejis.2013.19},
	doi = {10.1057/ejis.2013.19},
	language = {en},
	number = {6},
	urldate = {2025-05-08},
	journal = {European Journal of Information Systems},
	author = {Niehaves, Björn and Plattfaut, Ralf},
	month = nov,
	year = {2014},
	pages = {708--726},
}

@article{gell_patterns_2015,
	title = {Patterns of {Technology} {Use} {Among} {Older} {Adults} {With} and {Without} {Disabilities}},
	volume = {55},
	issn = {1758-5341, 0016-9013},
	url = {https://academic.oup.com/gerontologist/article/588723/Patterns},
	doi = {10.1093/geront/gnt166},
	language = {en},
	number = {3},
	urldate = {2025-05-08},
	journal = {The Gerontologist},
	author = {Gell, Nancy M. and Rosenberg, Dori E. and Demiris, George and LaCroix, Andrea Z. and Patel, Kushang V.},
	month = jun,
	year = {2015},
	pages = {412--421},
}

@article{lee_age_2011,
	title = {Age differences in constraints encountered by seniors in their use of computers and the internet},
	volume = {27},
	copyright = {https://www.elsevier.com/tdm/userlicense/1.0/},
	issn = {07475632},
	url = {https://linkinghub.elsevier.com/retrieve/pii/S0747563211000070},
	doi = {10.1016/j.chb.2011.01.003},
	language = {en},
	number = {3},
	urldate = {2025-05-08},
	journal = {Computers in Human Behavior},
	author = {Lee, Bob and Chen, Yiwei and Hewitt, Lynne},
	month = may,
	year = {2011},
	pages = {1231--1237},
}

@inproceedings{Zhao_2023, address={Hamburg Germany}, title={Older Adults Using Technology for Meaningful Activities During COVID-19: An Analysis Through the Lens of Self-Determination Theory}, ISBN={9781450394215}, url={https://dl.acm.org/doi/10.1145/3544548.3580839}, DOI={10.1145/3544548.3580839}, booktitle={Proceedings of the 2023 CHI Conference on Human Factors in Computing Systems}, publisher={ACM}, author={Zhao, Wei and Kelly, Ryan M. and Rogerson, Melissa J. and Waycott, Jenny}, year={2023}, month=apr, pages={1–17}, language={en} }

@inproceedings{Jelen2023,
author = {Jelen, Ben and Lazar, Amanda and Harrington, Christina and Pradhan, Alisha and Siek, Katie A.},
title = {Speaking from Experience: Co-designing E-textile Projects with Older Adult Fiber Crafters},
year = {2023},
isbn = {9781450399777},
publisher = {Association for Computing Machinery},
address = {New York, NY, USA},
url = {https://doi-org.proxy.library.nyu.edu/10.1145/3569009.3572736},
doi = {10.1145/3569009.3572736},
booktitle = {Proceedings of the Seventeenth International Conference on Tangible, Embedded, and Embodied Interaction},
articleno = {5},
numpages = {22},
location = {Warsaw, Poland},
series = {TEI '23}
}

@article{Var2026, title={Investigating Writing Professionals' Relationships with GenAI: How Combined Perceptions of Rivalry and Collaboration Shape Work Practices and Outcomes}, url={https://doi.org/10.1145/3772318.3790466}, DOI={10.1145/3772318.3790466}, author={Varanasi, Rama Adithya and Nov, Oded  and Wiesenfeld, Batia Mishan}, year={2026}, month=april }

@book{Hennink_2019, address={Thousand Oaks}, edition={2nd}, title={Qualitative research methods}, ISBN={9781473903913}, publisher={SAGE Publications Ltd}, author={Hennink, Monique and Hutter, Inge and Bailey, Ajay}, year={2019} }

@article{Braun_Clarke_2006, title={Using thematic analysis in psychology}, volume={3}, ISSN={1478-0887, 1478-0895}, url={http://www.tandfonline.com/doi/abs/10.1191/1478088706qp063oa}, DOI={10.1191/1478088706qp063oa}, number={2}, journal={Qualitative Research in Psychology}, author={Braun, Virginia and Clarke, Victoria}, year={2006}, month=jan, pages={77–101}, language={en} }

@inproceedings{Lee_2025, address={Yokohama Japan}, title={The Impact of Generative AI on Critical Thinking: Self-Reported Reductions in Cognitive Effort and Confidence Effects From a Survey of Knowledge Workers}, ISBN={9798400713941}, url={https://dl.acm.org/doi/10.1145/3706598.3713778}, DOI={10.1145/3706598.3713778}, booktitle={Proceedings of the 2025 CHI Conference on Human Factors in Computing Systems}, publisher={ACM}, author={Lee, Hao-Ping (Hank) and Sarkar, Advait and Tankelevitch, Lev and Drosos, Ian and Rintel, Sean and Banks, Richard and Wilson, Nicholas}, year={2025}, month=apr, pages={1–22}, language={en} }

@article{Convertino_2007, title={Supporting intergenerational groups in computer-supported cooperative work (CSCW)}, volume={26}, ISSN={0144-929X, 1362-3001}, url={http://www.tandfonline.com/doi/abs/10.1080/01449290601173473}, DOI={10.1080/01449290601173473}, number={4}, journal={Behaviour \& Information Technology}, author={Convertino, G. and Farooq, U. and Rosson, M. B. and Carroll, J. M. and Meyer, B. J. F.}, year={2007}, month=jul, pages={275–285}, language={en} }

@inproceedings{Kope2018,
author = {Kope\'{c}, Wies\l{}aw and Balcerzak, Bart\l{}omiej and Nielek, Radoslaw and Kowalik, Grzegorz and Wierzbicki, Adam and Casati, Fabio},
title = {Older adults and hackathons: a qualitative study},
year = {2018},
isbn = {9781450356381},
publisher = {Association for Computing Machinery},
address = {New York, NY, USA},
url = {https://doi-org.proxy.library.nyu.edu/10.1145/3180155.3182547},
doi = {10.1145/3180155.3182547},
booktitle = {Proceedings of the 40th International Conference on Software Engineering},
pages = {702–703},
numpages = {2},
location = {Gothenburg, Sweden},
series = {ICSE '18}
}

@inproceedings{pang_technology_2021,
	address = {Yokohama Japan},
	title = {Technology {Adoption} and {Learning} {Preferences} for {Older} {Adults}: {Evolving} {Perceptions}, {Ongoing} {Challenges}, and {Emerging} {Design} {Opportunities}},
	isbn = {978-1-4503-8096-6},
	shorttitle = {Technology {Adoption} and {Learning} {Preferences} for {Older} {Adults}},
	url = {https://dl.acm.org/doi/10.1145/3411764.3445702},
	doi = {10.1145/3411764.3445702},
	language = {en},
	urldate = {2025-05-03},
	booktitle = {Proceedings of the 2021 {CHI} {Conference} on {Human} {Factors} in {Computing} {Systems}},
	publisher = {ACM},
	author = {Pang, Carolyn and Collin Wang, Zhiqin and McGrenere, Joanna and Leung, Rock and Dai, Jiamin and Moffatt, Karyn},
	month = may,
	year = {2021},
	pages = {1--13},
}

@book{suchman2007human,
  title={Human-machine reconfigurations: Plans and situated actions},
  author={Suchman, Lucille Alice},
  year={2007},
  publisher={Cambridge university press}
}

@article{brynjolfsson_productivity_2021,
	title = {The {Productivity} {J}-{Curve}: {How} {Intangibles} {Complement} {General} {Purpose} {Technologies}},
	volume = {13},
	issn = {1945-7707, 1945-7715},
	shorttitle = {The {Productivity} {J}-{Curve}},
	url = {https://pubs.aeaweb.org/doi/10.1257/mac.20180386},
	doi = {10.1257/mac.20180386},
	language = {en},
	number = {1},
	urldate = {2025-05-08},
	journal = {American Economic Journal: Macroeconomics},
	author = {Brynjolfsson, Erik and Rock, Daniel and Syverson, Chad},
	month = jan,
	year = {2021},
	pages = {333--372},
}

@article{lee_perspective_2015,
	title = {{PERSPECTIVE}: {Older} {Adults}' {Adoption} of {Technology}: {An} {Integrated} {Approach} to {Identifying} {Determinants} and {Barriers}},
	volume = {32},
	copyright = {http://onlinelibrary.wiley.com/termsAndConditions\#vor},
	issn = {0737-6782, 1540-5885},
	shorttitle = {{PERSPECTIVE}},
	url = {https://onlinelibrary.wiley.com/doi/10.1111/jpim.12176},
	doi = {10.1111/jpim.12176},
	language = {en},
	number = {5},
	urldate = {2025-05-08},
	journal = {Journal of Product Innovation Management},
	author = {Lee, Chaiwoo and Coughlin, Joseph F.},
	month = sep,
	year = {2015},
	pages = {747--759},
}

@misc{czaja_impact_2007,
	title = {The impact of aging on access to technology},
	url = {https://colab.ws/articles/10.1145%2F1102187.1102189},
	language = {en},
	urldate = {2025-02-22},
	journal = {CoLab},
	author = {Czaja, Sara J.},
	year = {2007},
}

@book{Butler_2010, address={New York}, title={The Longevity Revolution: the Benefits and Challenges of Living a Long Life.}, ISBN={9781586488550}, abstractNote={Pulitzer-prize winning author Dr. Robert Butler coined the term “ageism” and made “Alzheimer’s” a familiar word. Now he brings his formidable knowledge and experience in aging issues to a recent and unprecedented achievement: the extension of human life expectancy by thirty years. As Butler shows, our society has not yet adapted to this change. The U.S. has not made a research investment in aging. Only eleven medical schools out of 145 have geriatrics departments compared to England where geriatrics is the number two specialty. We have not solidified private pension plans or strengthened Socia.}, publisher={Perseus Books Group}, author={Butler, Robert N.}, year={2010}, language={eng} }

@inproceedings{lazar_going_2017,
	address = {Portland Oregon USA},
	title = {Going {Gray}, {Failure} to {Hire}, and the {Ick} {Factor}: {Analyzing} {How} {Older} {Bloggers} {Talk} about {Ageism}},
	isbn = {978-1-4503-4335-0},
	shorttitle = {Going {Gray}, {Failure} to {Hire}, and the {Ick} {Factor}},
	url = {https://dl.acm.org/doi/10.1145/2998181.2998275},
	doi = {10.1145/2998181.2998275},
	language = {en},
	urldate = {2025-04-28},
	booktitle = {Proceedings of the 2017 {ACM} {Conference} on {Computer} {Supported} {Cooperative} {Work} and {Social} {Computing}},
	publisher = {ACM},
	author = {Lazar, Amanda and Diaz, Mark and Brewer, Robin and Kim, Chelsea and Piper, Anne Marie},
	month = feb,
	year = {2017},
	pages = {655--668},
}

@article{vines_age-old_2015,
	title = {An {Age}-{Old} {Problem}: {Examining} the {Discourses} of {Ageing} in {HCI} and {Strategies} for {Future} {Research}},
	volume = {22},
	issn = {1073-0516, 1557-7325},
	shorttitle = {An {Age}-{Old} {Problem}},
	url = {https://dl.acm.org/doi/10.1145/2696867},
	doi = {10.1145/2696867},
	language = {en},
	number = {1},
	urldate = {2025-04-28},
	journal = {ACM Transactions on Computer-Human Interaction},
	author = {Vines, John and Pritchard, Gary and Wright, Peter and Olivier, Patrick and Brittain, Katie},
	month = mar,
	year = {2015},
	pages = {1--27},
}

@article{Peng_Chan_2019, title={A meta-analysis of the relationship between ageing and occupational safety and health}, volume={112}, ISSN={09257535}, url={https://linkinghub.elsevier.com/retrieve/pii/S0925753518304685}, DOI={10.1016/j.ssci.2018.10.030}, journal={Safety Science}, author={Peng, Lu and Chan, Alan H.S.}, year={2019}, month=feb, pages={162–172}, language={en} }

@inproceedings{cuadra23,
author = {Cuadra, Andrea and Bethune, Jessica and Krell, Rony and Lempel, Alexa and H\"{a}nsel, Katrin and Shahrokni, Armin and Estrin, Deborah and Dell, Nicola},
title = {Designing Voice-First Ambient Interfaces to Support Aging in Place},
year = {2023},
isbn = {9781450398930},
publisher = {Association for Computing Machinery},
address = {New York, NY, USA},
url = {https://doi-org.proxy.library.nyu.edu/10.1145/3563657.3596104},
doi = {10.1145/3563657.3596104},
booktitle = {Proceedings of the 2023 ACM Designing Interactive Systems Conference},
pages = {2189–2205},
numpages = {17},
location = {Pittsburgh, PA, USA},
series = {DIS '23}
}

@article{Pradhan2020,
author = {Pradhan, Alisha and Lazar, Amanda and Findlater, Leah},
title = {Use of Intelligent Voice Assistants by Older Adults with Low Technology Use},
year = {2020},
issue_date = {August 2020},
publisher = {Association for Computing Machinery},
address = {New York, NY, USA},
volume = {27},
number = {4},
issn = {1073-0516},
url = {https://doi-org.proxy.library.nyu.edu/10.1145/3373759},
doi = {10.1145/3373759},
journal = {ACM Trans. Comput.-Hum. Interact.},
month = sep,
articleno = {31},
numpages = {27}
}

@article{jain_population_2025,
	title = {Population age structural transition, demographic dividend and economic growth in {India}},
	volume = {12},
	issn = {2662-9992},
	url = {https://www.nature.com/articles/s41599-025-05042-0},
	doi = {10.1057/s41599-025-05042-0},
	language = {en},
	number = {1},
	urldate = {2025-09-01},
	journal = {Humanities and Social Sciences Communications},
	author = {Jain, Neha and Goli, Srinivas and Jana, Arjun},
	month = jun,
	year = {2025},
	pages = {771},
}

@book{disalvo_adversarial_2015,
	address = {Cambridge, Massachusetts},
	series = {Design thinking, design theory},
	title = {Adversarial design},
	isbn = {978-0-262-52822-1},
	language = {eng},
	publisher = {The MIT Press},
	author = {DiSalvo, Carl},
	year = {2015},
}

@article{vollset2024burden,
  title={Burden of disease scenarios for 204 countries and territories, 2022--2050: a forecasting analysis for the Global Burden of Disease Study 2021},
  author={Vollset, Stein Emil and Ababneh, Hazim S and Abate, Yohannes Habtegiorgis and Abbafati, Cristiana and Abbasgholizadeh, Rouzbeh and Abbasian, Mohammadreza and Abbastabar, Hedayat and Abd Al Magied, Abdallah HA and Abd ElHafeez, Samar and Abdelkader, Atef and others},
  journal={The Lancet},
  volume={403},
  number={10440},
  pages={2204--2256},
ISSN={01406736}, 
url={https://linkinghub.elsevier.com/retrieve/pii/S0140673624006858}, 
DOI={10.1016/S0140-6736(24)00685-8}, 
  year={2024},
  publisher={Elsevier}
}

@article{frank2019toward,
  title={Toward understanding the impact of artificial intelligence on labor},
  author={Frank, Morgan R and Autor, David and Bessen, James E and Brynjolfsson, Erik and Cebrian, Manuel and Deming, David J and Feldman, Maryann and Groh, Matthew and Lobo, Jos{\'e} and Moro, Esteban and others},
  journal={Proceedings of the National Academy of Sciences},
  volume={116},
  number={14},
  pages={6531--6539},
  year={2019},
  publisher={National Academy of Sciences}
}

@misc{berger_defined_2012,
	title = {From {Defined} {Benefit} to {Defined} {Contribution}: {A} {Systematic} {Approach} to {Transitioning} {Retirement} {Plans}},
	shorttitle = {From {Defined} {Benefit} to {Defined} {Contribution}},
	url = {https://www.shrm.org/topics-tools/news/benefits-compensation/defined-benefit-to-defined-contribution-systematic-approach-to-transitioning-retirement-plans},
	language = {en-US},
	urldate = {2025-09-09},
	author = {Berger, Rich},
	month = jan,
	year = {2012},
}

@misc{path_changing_2024,
	title = {The {Changing} {Landscape} of {Retirement} in {India}},
	url = {https://wisdomcircle.com/changing-landscape-of-retirement-in-india/},
	language = {en-US},
	urldate = {2025-09-09},
	author = {{WisdomCircle}},
	month = apr,
	year = {2024}
}

@article{sahoo_charting_2024,
	title = {Charting the course: {India}’s health expenditure projections for 2035},
	volume = {8},
	issn = {2414-6447},
	shorttitle = {Charting the course},
	url = {https://www.sciencedirect.com/science/article/pii/S2414644724000228},
	doi = {10.1016/j.glohj.2024.05.001},
	number = {2},
	urldate = {2025-09-09},
	journal = {Global Health Journal},
	author = {Sahoo, Pragyan Monalisa and Rout, Himanshu Sekhar},
	month = jun,
	year = {2024},
	pages = {58--66},
}

@misc{inamdar_tcs_2025,
	title = {{TCS}: {India}'s {AI}-driven tech firings could derail middle class dreams},
	shorttitle = {{TCS}},
	url = {https://www.bbc.com/news/articles/cx2p4nqd352o},
	language = {en-GB},
	urldate = {2025-09-09},
	author = {Inamdar, Nikhil},
	month = jul,
	year = {2025},
}

@misc{rajmohan_impact_2025,
	title = {Impact of {AI} on {White} {Collar} {Indian} {Work}: {An} {Analysis} of ‘{Bullshit} {Jobs}’ {\textbar} {TechPolicy}.{Press}},
	shorttitle = {Impact of {AI} on {White} {Collar} {Indian} {Work}},
	url = {https://techpolicy.press/impact-of-ai-on-white-collar-indian-work-an-analysis-of-bullshit-jobs},
	language = {en},
	urldate = {2025-09-09},
	journal = {Tech Policy Press},
	author = {Rajmohan, Karthika},
	month = mar,
	year = {2025},
}

@techreport{chakrabarti_labour-force_2024,
	title = {Labour-force {Perception} about {AI}: {A} {Study} on {Indian} {White}-collar {Workers}},
	institution = {Brij Disa Centre for Data SCience and Artificial Intelligence, IIM Ahmedabad},
	author = {Chakrabarti, Anindya and Ghatak, Debjit and Moses, Aditya C. and Mukherjee, Deep Narayan and Sinha, Ankur and Todkar, Amita},
	month = aug,
	year = {2024},
}

@misc{indiaai_india_2026,
	title = {India {Leads} in {AI} {Adoption}, {Says} {BCG} {Study}},
	url = {https://indiaai.gov.in/news/india-leads-in-ai-adoption-says-bcg-study},
	language = {en},
	urldate = {2025-09-09},
	journal = {IndiaAI},
	author = {IndiaAI},
	month = nov,
	year = {2026},
}

@book{Gieryn_1999, address={Chicago}, title={Cultural Boundaries of Science: Credibility on the Line}, ISBN={9780226292618}, url = {https://doi.org/10.7208/chicago/9780226824420}, publisher={University of Chicago Press}, author={Gieryn, Thomas F.}, year={1999}, language={eng} }

@article{Farchi2023, title={Do We Still Need Professional Boundaries? The multiple influences of boundaries on interprofessional collaboration}, volume={44}, ISSN={0170-8406, 1741-3044}, url={https://journals.sagepub.com/doi/10.1177/01708406221074146}, DOI={10.1177/01708406221074146},  number={2}, journal={Organization Studies}, author={Farchi, Tomas and Dopson, Sue and Ferlie, Ewan}, year={2023}, month=feb, pages={277–298}, language={en} }

@article{zhao_employees_2024,
	title = {Employees’ perception of generative artificial intelligence and the dark side of work outcomes},
	volume = {61},
	issn = {14476770},
	url = {https://linkinghub.elsevier.com/retrieve/pii/S1447677024001207},
	doi = {10.1016/j.jhtm.2024.10.007},
	language = {en},
	urldate = {2025-09-10},
	journal = {Journal of Hospitality and Tourism Management},
	author = {Zhao, Hairong and Yuan, Bocong and Song, Yang},
	month = dec,
	year = {2024},
	pages = {191--199},
}

@inproceedings{shukla_-skilling_2025,
	address = {Yokohama Japan},
	title = {De-skilling, {Cognitive} {Offloading}, and {Misplaced} {Responsibilities}: {Potential} {Ironies} of {AI}-{Assisted} {Design}},
	isbn = {979-8-4007-1395-8},
	shorttitle = {De-skilling, {Cognitive} {Offloading}, and {Misplaced} {Responsibilities}},
	url = {https://dl.acm.org/doi/10.1145/3706599.3719931},
	doi = {10.1145/3706599.3719931},
	language = {en},
	urldate = {2025-09-10},
	booktitle = {Proceedings of the {Extended} {Abstracts} of the {CHI} {Conference} on {Human} {Factors} in {Computing} {Systems}},
	publisher = {ACM},
	author = {Shukla, Prakash and Bui, Phuong and Levy, Sean S and Kowalski, Max and Baigelenov, Ali and Parsons, Paul},
	month = apr,
	year = {2025},
	pages = {1--7},
}

@phdthesis{saradha_vathana_selvendran_bridge_2022,
	title = {Bridge {Employment}: {A} {Systematic} {Literature} {Review} on the {Association} {Between} {Bridge} {Employment} and {Working} {Retirees}' {Well}-{Being} and {Mental} health},
	school = {University of Groningen},
	author = {{Saradha Vathana Selvendran}},
	month = feb,
	year = {2022},
}

@article{oh_bridge_2024,
	title = {Bridge {Employment} or {Encore} {Career}? {Examining} {Predictors} {That} {Distinguish} {Later}-{Life} {Career} {Transitions}},
	volume = {79},
	copyright = {https://academic.oup.com/pages/standard-publication-reuse-rights},
	issn = {1079-5014, 1758-5368},
	shorttitle = {Bridge {Employment} or {Encore} {Career}?},
	url = {https://academic.oup.com/psychsocgerontology/article/doi/10.1093/geronb/gbae104/7688323},
	doi = {10.1093/geronb/gbae104},
	language = {en},
	number = {8},
	urldate = {2025-09-11},
	journal = {The Journals of Gerontology, Series B: Psychological Sciences and Social Sciences},
	author = {Oh, Yun Taek},
	editor = {Schafer, Markus},
	month = aug,
	year = {2024},
	pages = {gbae104},
}

@article{massihzadegan_preparing_2022,
	title = "Preparing Older Adults for Remote Employment: Opportunities and Challenges",
	volume = {6},
	copyright = {https://creativecommons.org/licenses/by/4.0/},
	issn = {2399-5300},
	url = {https://academic.oup.com/innovateage/article/6/Supplement_1/676/6938281},
	doi = {10.1093/geroni/igac059.2487},
	language = {en},
	number = {Supplement\_1},
	urldate = {2025-09-11},
	journal = {Innovation in Aging},
	author = {Massihzadegan, Setarreh and Gleason, Shayna and Mutchler, Jan and Coyle, Caitlin},
	month = dec,
	year = {2022},
	pages = {676--676}
}

@article{sharit_employability_2009,
	title = {The employability of older workers as teleworkers: {An} appraisal of issues and an empirical study},
	volume = {19},
	copyright = {http://onlinelibrary.wiley.com/termsAndConditions\#vor},
	issn = {1090-8471, 1520-6564},
	shorttitle = {The employability of older workers as teleworkers},
	url = {https://onlinelibrary.wiley.com/doi/10.1002/hfm.20138},
	doi = {10.1002/hfm.20138},
	language = {en},
	number = {5},
	urldate = {2025-09-11},
	journal = {Human Factors and Ergonomics in Manufacturing \& Service Industries},
	author = {Sharit, Joseph and Czaja, Sara J. and Hernandez, Mario A. and Nair, Sankaran N.},
	month = sep,
	year = {2009},
	pages = {457--477},
}

@article{aisa_automation_2023,
	title = {Automation and aging: {The} impact on older workers in the workforce},
	volume = {26},
	issn = {2212828X},
	shorttitle = {Automation and aging},
	url = {https://linkinghub.elsevier.com/retrieve/pii/S2212828X23000361},
	doi = {10.1016/j.jeoa.2023.100476},
	language = {en},
	urldate = {2025-09-11},
	journal = {The Journal of the Economics of Ageing},
	author = {Aisa, Rosa and Cabeza, Josefina and Martin, Jorge},
	month = oct,
	year = {2023},
	pages = {100476},
}

@article{dutta_elderly_2026,
	title = {Elderly welfare in {India}: {Navigating} challenges and inclusive social interventions},
	volume = {8},
	issn = {27099350, 27099369},
	shorttitle = {Elderly welfare in {India}},
	url = {https://www.multisubjectjournal.com/archives/2026.v8.i3.A.932},
	doi = {10.22271/multi.2026.v8.i3a.932},
	language = {en},
	number = {3},
	urldate = {2026-04-13},
	journal = {International Journal of Multidisciplinary Trends},
	author = {Dutta, Manaswita and Singh, Nandini C and Wangsa, Sainon J and Bhujel, Suman},
	month = mar,
	year = {2026},
	pages = {05--09},
}

@techreport{helpage_india_ageing_2024,
	title = {Ageing in {India}: {Exploring} {Preparedness} {Response} to {Care} {Challenges}},
	institution = {Helpage India},
	author = {Helpage India},
	year = {2024},
}

@misc{mahambare_india_2026,
	title = {India is facing a silent elderly care crisis. {Budget} 2026 must confront it},
	url = {https://ncaer.org/publication/india-is-facing-a-silent-elderly-care-crisis-budget-2026-must-confront-it/},
	language = {en-US},
	urldate = {2026-04-13},
	journal = {NCAER {\textbar} Quality . Relevance . Impact},
	author = {Mahambare, Vidya and Munjal, Poonam and Baruah, Palash},
	year = {2026},
}

@misc{chakrabarty_union_2026,
	title = {Union {Budget} 2026-27: {Neglecting} {India}’s {Elders}},
	shorttitle = {Union {Budget} 2026-27},
	url = {https://www.orfonline.org/expert-speak/union-budget-2026-27-neglecting-india-s-elders},
	language = {en},
	urldate = {2026-04-13},
	journal = {orfonline.org},
	publisher = {Observer Research Foundation},
	author = {Chakrabarty, Malancha},
	year = {2026},
}

@misc{iyr_hidden_2025,
	title = {The {Hidden} {Cost} {Of} {India}’s {Economic} {Boom}: {Inequality} {At} {Historic} {High}, {As} {Wealth} \& {Income} {Growth} {Leave} {Out} {Millions}},
	shorttitle = {The {Hidden} {Cost} {Of} {India}’s {Economic} {Boom}},
	url = {https://article-14.com/post/the-hidden-cost-of-india-s-economic-boom-inequality-at-historic-high-as-wealth-income-growth-leave-out-millions--693b8b0d6e810},
	urldate = {2026-04-13},
	publisher = {Article 14},
	author = {Iyr, Kavitha},
	month = dec,
	year = {2025},
}

@article{van2013ability,
  title={The ability of older people to overcome adversity: A review of the resilience concept},
  author={Van Kessel, Gisela},
  journal={Geriatric nursing},
  volume={34},
  number={2},
  pages={122--127},
  year={2013},
  publisher={Elsevier}
}

@article{vyas2017everyday,
  title={Everyday resilience: Supporting resilient strategies among low socioeconomic status communities},
  author={Vyas, Dhaval and Dillahunt, Tawanna},
  journal={Proceedings of the ACM on Human-Computer Interaction},
  volume={1},
  number={CSCW},
  pages={1--21},
  year={2017},
  publisher={ACM New York, NY, USA}
}

@article{ye_psychological_2024,
	title = {Psychological {Resilience} and {Frailty} {Progression} in {Older} {Adults}},
	volume = {7},
	issn = {2574-3805},
	url = {https://jamanetwork.com/journals/jamanetworkopen/fullarticle/2827022},
	doi = {10.1001/jamanetworkopen.2024.47605},
	language = {en},
	number = {11},
	urldate = {2026-04-15},
	journal = {JAMA Network Open},
	author = {Ye, Bo and Li, Yunxia and Bao, Zhijun and Gao, Junling},
	month = nov,
	year = {2024},
	pages = {e2447605},
}

@article{lima_resilience_2023,
	title = {Resilience in {Older} {People}: {A} {Concept} {Analysis}},
	volume = {11},
	issn = {2227-9032},
	shorttitle = {Resilience in {Older} {People}},
	url = {https://www.mdpi.com/2227-9032/11/18/2491},
	doi = {10.3390/healthcare11182491},
	language = {en},
	number = {18},
	urldate = {2026-04-15},
	journal = {Healthcare},
	author = {Lima, Gabriella Santos and Figueira, Ana Laura Galhardo and Carvalho, Emília Campos De and Kusumota, Luciana and Caldeira, Sílvia},
	month = sep,
	year = {2023},
	pages = {2491},
}

@article{cosarderelioglu_frailty_2025,
	title = {From frailty to resilience: exploring adaptive capacity and reserve in older adults–a narrative review},
	volume = {6},
	issn = {2673-6217},
	shorttitle = {From frailty to resilience},
	url = {https://www.frontiersin.org/articles/10.3389/fragi.2025.1520842/full},
	doi = {10.3389/fragi.2025.1520842},
	urldate = {2026-04-15},
	journal = {Frontiers in Aging},
	author = {Cosarderelioglu, Caglar and Walston, Jeremy D. and Abadir, Peter M.},
	month = jul,
	year = {2025},
	pages = {1520842},
}

@article{guan_research_2025,
	title = {Research on the impact of delayed retirement on the subjective wellbeing of older adults},
	volume = {13},
	issn = {2296-2565},
	url = {https://www.frontiersin.org/articles/10.3389/fpubh.2025.1530613/full},
	doi = {10.3389/fpubh.2025.1530613},
	urldate = {2026-04-15},
	journal = {Frontiers in Public Health},
	author = {Guan, Guo-feng and Wang, Lu and Ma, Dong-fen and Luo, Pan-shi},
	month = jun,
	year = {2025},
	pages = {1530613},
}

@misc{money_control_low_2025,
	title = {From low pensions to delayed dues: {These} retirement disputes could affect you, too- {Moneycontrol}.com},
	shorttitle = {From low pensions to delayed dues},
	url = {https://www.moneycontrol.com/news/business/personal-finance/from-low-pensions-to-delayed-dues-these-retirement-disputes-could-affect-you-too-13721718.html},
	language = {en},
	urldate = {2026-04-15},
	journal = {Moneycontrol},
	author = {Money Control},
	month = dec,
	year = {2025},
}

@misc{sun_life_news_2026,
	title = {Sun {Life} {Survey} {Reveals} {Retirement} {Divide} in {Asia}: {An} {Option} for {Some}, an {Obligation} for {Others}},
	url = {https://www.sunlife.com/en/newsroom/news-releases/announcement/sun-life-survey-reveals-retirement-divide-in-asia-an-option-for-some-an-obligation-for-others/124053/},
	language = {en},
	urldate = {2026-04-15},
	author = {{Sun Life}},
	month = jan,
	year = {2026},
}

@misc{bungsut_age_2024,
	title = {Age against the machine: {The} need for a pension revolution in {India}},
	shorttitle = {Age against the machine},
	url = {https://www.developmentpathways.co.uk/blog/age-against-the-machine-the-need-for-a-pension-revolution-in-india/},
	language = {en-GB},
	urldate = {2026-04-15},
	journal = {Development Pathways},
	author = {Bungsut, Chhani},
	month = oct,
	year = {2024},
}

@book{wenger_communities_1998,
	edition = {1},
	title = {Communities of {Practice}: {Learning}, {Meaning}, and {Identity}},
	copyright = {https://www.cambridge.org/core/terms},
	isbn = {978-0-521-43017-3 978-0-521-66363-2 978-0-511-80393-2},
	shorttitle = {Communities of {Practice}},
	url = {https://www.cambridge.org/core/product/identifier/9780511803932/type/book},
	doi = {10.1017/CBO9780511803932},
	urldate = {2026-04-15},
	publisher = {Cambridge University Press},
	author = {Wenger, Etienne},
	month = jul,
	year = {1998},
}

@article{brown_organizational_1991,
	title = {Organizational {Learning} and {Communities}-of-{Practice}: {Toward} a {Unified} {View} of {Working}, {Learning}, and {Innovation}},
	volume = {2},
	issn = {1047-7039, 1526-5455},
	shorttitle = {Organizational {Learning} and {Communities}-of-{Practice}},
	url = {https://pubsonline.informs.org/doi/10.1287/orsc.2.1.40},
	doi = {10.1287/orsc.2.1.40},
	language = {en},
	number = {1},
	urldate = {2026-04-15},
	journal = {Organization Science},
	author = {Brown, John Seely and Duguid, Paul},
	month = feb,
	year = {1991},
	pages = {40--57},
}

@inproceedings{millen_improving_2003,
	address = {Sanibel Island Florida USA},
	title = {Improving individual and organizational performance through communities of practice},
	isbn = {978-1-58113-693-7},
	url = {https://dl.acm.org/doi/10.1145/958160.958192},
	doi = {10.1145/958160.958192},
	language = {en},
	urldate = {2026-04-15},
	booktitle = {Proceedings of the 2003 international {ACM} {SIGGROUP} conference on {Supporting} group work},
	publisher = {ACM},
	author = {Millen, David R. and Fontaine, Michael A.},
	month = nov,
	year = {2003},
	pages = {205--211},
}

\end{document}